%% file: 41619corr.tex
\documentclass[onecolumn]{aa}

\usepackage[english]{babel}
\usepackage{MnSymbol}
\usepackage{tikz}
\usetikzlibrary{shapes.geometric, arrows}
\usepackage{natbib}
\bibpunct{(}{)}{;}{a}{}{,}
\usepackage[breaklinks, colorlinks, citecolor=blue]{hyperref}

\input{Planck.tex}

\input{shortcuts.tex}

\begin{document}
	
	\title{Detailed study of HWP non-idealities and their impact on future measurements of CMB polarization anisotropies from space}

	\author{S.~Giardiello \inst{1,2}~\thanks{Corresponding author: Serena~Giardiello, serena.giardiello@unife.it}  \and M.~Gerbino \inst{2} \and L.~Pagano \inst{1,2} \and J.~Errard \inst{3} \and A.~Gruppuso \inst{4,5} \and H.~Ishino \inst{6} \and M.~Lattanzi \inst{2} \and P.~Natoli \inst{1,2} \and G.~Patanchon \inst{3} \and F.~Piacentini \inst{7,8} \and G.~Pisano \inst{7}}

	\institute{Dipartimento di Fisica e Scienze della Terra, Universit\`a di Ferrara,     Polo Scientifico e Tecnologico - Edificio C Via Saragat, 1, I-44122, Ferrara, Italy \and Istituto Nazionale di Fisica Nucleare, sezione di Ferrara,     Polo Scientifico e Tecnologico - Edificio C Via Saragat, 1, I-44122, Ferrara, Italy \and Universit\'e de Paris, CNRS, Astroparticule et Cosmologie,
		F-75006 Paris, France \and Istituto Nazionale di Astrofisica, Osservatorio di Astrofisica e Scienza
		dello Spazio di Bologna, via Gobetti 101, I-40129 Bologna, Italy
		\and Istituto Nazionale di Fisica Nucleare, Sezione di Bologna, viale Berti
		Pichat 6/2, I-40127 Bologna, Italy \and Department of Physics, Okayama University, 3-1-1 Tsushimanaka,
		Kita-ku, Okayama 700-8530, Japan \and Dipartimento di Fisica, Sapienza Universit\`a di Roma, sezione di Roma1, P.le A. Moro
		5, 00185, Roma, Italy \and Istituto Nazionale di Fisica Nucleare, sezione di Roma1, P.le A. Moro
		5, 00185, Roma, Italy }

	\titlerunning{HWP non-idealities: a detailed study}
	\authorrunning{Giardiello et al.}

	\abstract{We study the propagation of a specific class of instrumental systematics to the reconstruction of the B-mode power spectrum of the cosmic microwave background (CMB). We focus on the non-idealities of the half-wave plate (HWP), a polarization modulator that is to be deployed by future CMB experiments, such as the phase-A satellite mission LiteBIRD. We study the effects of non-ideal HWP properties, such as transmittance, phase shift, and cross-polarization. To this end, we developed a simple, yet stand-alone end-to-end simulation pipeline adapted to LiteBIRD. We analyzed the effects of a possible mismatch between the measured frequency profiles of HWP properties (used in the mapmaking stage of the pipeline) and the actual profiles (used in the sky-scanning step). We simulated single-frequency, CMB-only observations to emphasize the effects of non-idealities on the BB power spectrum. We also considered multi-frequency observations to account for the frequency dependence of HWP properties and the contribution of foreground emission. We quantified the systematic effects in terms of a bias $\Delta r$ on the tensor-to-scalar ratio, $r,$ with respect to the ideal case without systematic effects. We derived the accuracy requirements on the measurements of HWP properties by requiring $\Delta r < 10^{-5}$ (1\% of the expected LiteBIRD sensitivity on $r$). Our analysis is introduced by a detailed presentation of the mathematical formalism employed in this work, including 
		the use of the Jones and Mueller matrix representations.
	}
	
	
	\maketitle
	
	\section{Introduction}\label{sec:intro}
	In past decades, the cosmic microwave background (CMB) has played a fundamental role in helping to improve our knowledge of the Universe. Measurements of CMB anisotropies in temperature and polarization (E-modes and lensing-induced B-modes) have been decisive in shaping the current cosmological model, from the quantum
	mechanical origin of the Universe, to its current energy composition \citep{hinshaw2012,planck2016-l01,Ade:2018,Aiola_2020,Bianchini_2020,P_A_R_Ade_2014}. Major advances in the observation of the polarized CMB signal are expected from the forthcoming generation of CMB experiments such as the ground-based Simons Observatory (SO) \citep{SO} and CMB-S4 \citep{Abazajian:2020dmr} as well as the LiteBIRD satellite mission \citep{Hazumi:2021yqq}. The most ambitious target is the measurement of the primordial B-mode signal. A high-significance detection of the latter would constrain the amplitude of primordial gravitational waves, parameterized in terms of the tensor-to-scalar ratio $r$. A combination of 
	state-of-the-art cosmological data \citep{Ade:2018gkx} provides the upper bound, $r<0.06$ at 95\% C.L., updated to $r<0.044$ (95\% CL) based on a recent re-analysis of Planck data~\citep{Tristram:2020wbi}. The detection of $r$ would strongly support the validity of the inflation paradigm. On the other hand, a tighter upper bound for $r<0.001$ would rule out a large proportion of consistently viable early Universe models (single-field models with typical inflaton excursion that is much larger than the Planck mass scale). 
	
	The ambitious sensitivity goals of future surveys, namely: $\sigma(r)\simeq0.002$ from SO, $r<0.001$ at 95\% CL from CMB-S4, and $\sigma(r)\simeq0.001$ from LiteBIRD, require extraordinary control over systematic effects and noise contamination \citep[LiteBIRD collaboration, in prep.;][]{SO}. 
	To that end, the use of a polarization modulator like a half-wave plate (HWP) has been included in the design of future surveys, from the Small Aperture Telescopes (SAT) of SO \citep{SO} to LiteBIRD \citep{Hazumi:2021yqq} and LSPE \citep{lspe2012}. The HWPs have been already deployed in many polarization-sensitive experiments: MAXIPOL \citep{Johnson:2006jk}, SPIDER \citep{Rahlin:2014rja}, ABS \citep{Kusaka:2013pla}, POLARBEAR  \citep{Hill:2016jhd},  PILOT \citep{Misawa:2014hka}, BLAST \citep{2016SPIE.9914E..0JG}, and EBEX \citep{ReichbornKjennerud:2010ja}.
	These experiments have shown that the use of a HWP can reduce both the $1/f$ noise~\citep[in the case of countinuous spinning;][]{Johnson:2006jk}
	and systematic uncertainties related to the pair differencing 
	of orthogonal detectors~\citep{Bryan_2016,Essinger_Hileman_2016}.
	However, pernicious systematic effects induced by non-idealities in manufactured HWPs can propagate through the analysis pipeline and bias the final estimation of cosmological parameters, including $r$. Therefore, a study of the impact of HWP non-idealities on high-level science products is required.
	
	The aim of this work is to provide an exhaustive summary of the mathematical formalism that fully characterizes the behavior of a non-ideal HWP in the context of CMB measurements. We applied this formalism to simulate the effect of HWP non-idealities on the observation of the full sky, using a LiteBIRD-like strategy~\citep{Hazumi:2021yqq}. A simple analysis is performed for single-frequency observations to showcase the effect of the different HWP systematic parameters at the power-spectrum level. We also conducted a more realistic, multi-frequency analysis where the impact of HWP non-idealities is quantified in terms of a bias in the determination of the tensor-to-scalar ratio, $r$.
	
	This paper is organized as follows. In Sect.~\ref{sec:formalism}, we lay down the mathematical formalism employed in our analysis. Particular care is devoted to clarify a common misunderstanding when dealing with the choice of the matrix formulation (Jones and Mueller) to describe propagation of light through optical systems. In Sect.~\ref{sec:strategy}, we describe the scanning strategy and map-making procedure adopted in our simulations. In Sect.~\ref{sec:mono}, we present a simple monochromatic analysis. In Sect.~\ref{sec:cromo}, we present the multi-frequency study and the requirements we set on each non-ideal parameter (summarized in Table \ref{tab:dr_dsyst}) to keep the bias on $r$ under a pre-defined threshold ($\Delta r \leq 10^{-5}$). Our conclusions are presented in Sect.~\ref{sec:conclusion}.

	\section{Matrix representation of HWP optical effects}\label{sec:formalism}
	In this section, we review the two main mathematical formalisms employed to characterize the optical effect of a HWP on incident radiation, namely, on the Jones and Mueller matrix formalisms. 
	
	First, we begin with some basic assumptions. Supposing that a quasi-monochromatic wave propagates along a direction orthogonal to the surface of an optical device, we define a coordinate system $x-y$ on the surface of the optical device, so that the incoming wave can be decomposed into an $x$-component, $E_x$ and a $y$-component $E_y$. A wave plate (or retarder) is a phase-shifter, that is, a non-depolarizing linear optical device that modifies the phase of the incident wave. An ideal HWP induces a phase shift of $\pi$ between the two orthogonal components, $E_{x,y}$, of the incident wave. The phase-shift is due to the fact that the components of the incident wave propagate through the HWP with a different index of refraction. The physical properties \citep[e.g., thickness of the plate in case of HWP made of birefringent crystal, design of the stack of mesh filters in case of mesh-HWPs;][]{Pisano:crosspolcite} 
	of the HWP can be tuned at the manufacturing stage in such a way that the difference between the optical paths of the two components of the incident wave result in a phase shift of $\pi$ once the signal emerges from the HWP.
	The optical axis of the HWP with the highest (or lowest) index of refraction is called the ``slow'' (or ``fast'') axis. 
	
	The linear response of the HWP to the incoming signal allows us to represent the output signal emerging from it via linear algebra, that is, via a simple matrix transformation of the input signal.
	The non-depolarizing property means that the HWP does not decorrelate or randomize the amplitude and phase of the orthogonal components of the incident wave. The non-depolarizing nature of the HWP allows use of the Jones matrix formalism as the matrix representation of the HWP. We will see later that a Mueller matrix approach is also allowed and entirely equivalent to the Jones formalism in this case (i.e., a non-depolarizing device). In the following, we make use of the Jones formalism to provide a much clearer description of the physical effects of HWP non-idealities. The Mueller formalism will be handy for the application of our analysis to future CMB missions. 
	We would like to stress that the choice of the matrix representation of the optical element is independent from the polarization state of the incoming signal. Whether or not an optical element can be represented in terms of a Jones matrix does only depend on the nature of the optical system. In particular, the choice of the matrix representation stems from the non-depolarizing nature of the device. It can be proven \citep[Sect. 2.10]{ellipsometry} 
	that for a non-depolarizing device, the degree of polarization of the outgoing signal is always greater than or equal to the degree of polarization of the incoming signal. In contrast, a depolarizing device transfers power out of polarized states into unpolarized states. As such, the degree of polarization of the signal coming out from a depolarizer can be lower than the degree of polarization of the incoming signal. This is the only effect that the Jones formalism is unable to capture. When dealing with such devices, it is better to rely on alternative formalisms. All other non-depolarizing optical systems, including the HWP, can be adequately represented with Jones matrices \citep{ellipsometry}.
	
	\subsection{Jones matrix formalism}
	The Jones matrix of an optical system, including that of a HWP, is a $2\times2$ complex matrix applicable to the $(E_x,E_y)$ Jones vector. It is fully characterized by seven real 
	parameters: the real and imaginary part of each matrix element, minus a global phase that is not measurable. 
	%
	The Jones matrix of an ideal HWP with fast axis either along the $x$-axis or $y$-axis is expressed as:
	\begin{equation}\label{eq:ideal}
		J_{HWP, \, id} \equiv  \begin{pmatrix}
			1 & 0\\
			0 & e^{\imath\pi}\\
		\end{pmatrix}=\begin{pmatrix}
			1 & 0\\
			0 & -1\\
		\end{pmatrix}.
	\end{equation}
	
	\noindent Equation \ref{eq:ideal} has a straightforward  interpretation: the field along $x$ is left unchanged by the optical element while the phase of the $y$-component is shifted by $\pi$. However, the behavior of a real HWP can deviate from the ideal case. An expression that also accounts for small deviations of the HWP matrix elements from the ideal case (systematic effects) is as follows~\citep{ODea:2006tvb}: 
	
	\begin{equation} \label{eq:realistic}
		J_{HWP} = \begin{pmatrix}
			1+h_{1} & \zeta_{1} e^{i \chi_1}\\
			\zeta_{2} e^{i \chi_2}& -(1+h_{2}) e^{i \beta} \\ 
		\end{pmatrix} \equiv \begin{pmatrix}
			A_1 & B_1 \\
			B_2 & A_2 \\
		\end{pmatrix}, 
	\end{equation}
	where $A_{1}$ is real and $A_{2}$, $B_{1,2}$ are complex numbers. The meaning of these non-ideal parameters is as follows: 
	        \begin{itemize}
	                \item[$\bullet$]  $h_{1}$ and $h_{2}$:  loss parameters describing the deviation from the unitary transmission of light components $E_{x}$, $E_{y}$. They are negatively defined parameters, with  a range of [-1,0] (light absorption + reflection). In the ideal case, $h_1=h_2=0$;
	                \item[$\bullet$] $\beta = \phi - \pi $: where $\phi$ is the phase shift between the two directions. It accounts for variations of the phase difference between $E_{x}$ and $E_{y}$ with respect to the nominal value of $\pi$ for an ideal HWP. In the ideal case, $\beta=0$;
	                \item[$\bullet$] $\zeta_{1,2}$ and $\chi_{1,2}$:  amplitudes and phases of the off-diagonal terms, coupling $E_{x}$ and $E_{y}$. In practice, if the incoming wave is fully polarized along $x$ ($y$), a spurious $y$ ($x$) component would show up in the outgoing wave. Hereafter, we refer to this effect as ``cross-polarization,''  In the ideal case, $\zeta_{1,2}=\chi_{1,2}=0$.
	        \end{itemize}
	
%
%
	
	So far, we have omitted the dependence of the HWP Jones matrix elements on the frequency of the incident wave.  The manufacturing of a HWP is always tuned such that a phase shift of $\pi$ between orthogonal components is realized at a given frequency. Therefore, the matrix elements in both Eq.~\ref{eq:ideal} and Eq.~\ref{eq:realistic} 
	are function of the incident frequency. We assess in Sect.~\ref{sec:cromo} the relevance of this aspect in the context of CMB observations.
	We have also omitted the dependance on the incident angle, which we neglect in this study (see Sect.~\ref{sec:Muell_formalism}). 
	
	In our analysis, we are interested in the possibility that a rotating HWP is employed to modulate the polarization signal. When the HWP is rotated by $\theta(t) \equiv  \omega t$, where $ \omega t$ is the (time-dependent) angle between the HWP fast axis and the $x$-axis and $\omega$ is the angular velocity of the HWP, the Jones matrix is transformed as follows: 
	
	\begin{equation}\label{eq:rotatedJones}
		\begin{split}
			&J_{RHWP}(\theta) = R^{T}(\theta)J_{HWP}R(\theta) = \begin{pmatrix}
				J_{11}(\theta) & J_{12}(\theta)\\
				J_{21}(\theta) & J_{22}(\theta)\\
			\end{pmatrix} , \quad \text{with} \quad R(\theta) =  \begin{pmatrix}
				\text{cos}\theta & \text{sin}\theta\\
				- \text{sin}\theta & \text{cos}\theta \\
			\end{pmatrix} ,
		\end{split}
	\end{equation}
	
	\noindent where the time dependence is understood. The explicit expressions of the matrix elements of $J_{RHWP}$ are:
	\begin{equation} \label{eq:rotatedJelements}
		\begin{split}
			J_{11}(\theta) &= (1+h_1)\cos^2\theta - (1+h_2) e^{i\beta} \sin^2\theta  - (\zeta_1e^{i\chi_1}+\zeta_2e^{i\chi_2}) \cos\theta \sin\theta  \equiv A_1 \cos^2\theta + A_2 \sin^2\theta - (B_1 + B_2) \cos\theta \sin\theta, \\
			J_{12}(\theta) &=\left[(1+h_1) + (1+h_2) e^{i\beta}\right] \cos\theta\sin\theta + \zeta_1e^{i\chi_1} \cos^2\theta - \zeta_2e^{i\chi_2}  \sin^2\theta \equiv\left[A_1 - A_2 \right] \cos\theta\sin\theta + B_1 \cos^2\theta - B_2  \sin^2\theta, \\
			J_{21}(\theta) &=\left[(1+h_1) + (1+h_2) e^{i\beta}\right] \cos\theta\sin\theta + \zeta_2e^{i\chi_2} \cos^2\theta - \zeta_1e^{i\chi_1}  \sin^2\theta \equiv\left[A_1 - A_2 \right] \cos\theta\sin\theta + B_2 \cos^2\theta - B_1  \sin^2\theta,\\
			J_{22}(\theta) &= (1+h_1)\sin^2\theta - (1+h_2) e^{i\beta} \cos^2\theta + (\zeta_1e^{i\chi_1}+\zeta_2e^{i\chi_2}) \cos\theta \sin\theta \equiv A_1 \sin^2\theta + A_2 \cos^2\theta + (B_1 + B_2) \cos\theta \sin\theta.
		\end{split}
	\end{equation}

	\noindent In the ideal case, Eq.~\ref{eq:rotatedJelements} is reduced to
	
	\begin{equation}
		J_{RHWP}^\mathrm{ideal}(\theta) = \begin{pmatrix}
			\cos 2\theta & \sin 2\theta\\
			\sin 2\theta & -\cos 2\theta\\
		\end{pmatrix}.
	\end{equation}
	
	The matrix $J_{RHWP}(\theta)$ is in the reference frame of the telescope. If we refer instead to a fixed reference frame on the sky, we have to take into account also the instrument orientation angle, $\psi$, such that the expression for the rotated matrix becomes~\citep{Bryan2010}:
	\begin{equation}\label{eq:finalJ_compl}
		J_{RHWP}(\theta) \rightarrow J_{RHWP}(\theta) R(\psi),
	\end{equation}
	which is equivalent to Eq.~\ref{eq:rotatedJones} but including the substitution: 
	\begin{equation}\label{eq:rotangle}
		\theta(t) = \omega t + \frac{\psi(t)}{2} .
	\end{equation}

	In this work, we need to take into account that the signal modulated by a rotating HWP is then collected by a polarization-sensitive detector. We consider pairs of polarization-sensitive detectors with orthogonal orientations, as those usually employed in CMB experiments in order to reconstruct the input sky signal more efficiently. The full optical chain traversed by incoming light that is perpendicular to the HWP is described via:
	\begin{equation}\label{eq:finalJ}
		\begin{split}
			&J_{\mathrm{tot},(x,y)}(\theta) = J_{pol,(x,y)}J_{RHWP}(\theta), \quad \text{where} \quad  J_{pol,x} = \begin{pmatrix} 
				1 & 0\\
				0 & 0\\
			\end{pmatrix} \quad \text{or} \quad  J_{pol,y} = \begin{pmatrix}
				0 & 0\\
				0 & 1\\
			\end{pmatrix}.
		\end{split}
	\end{equation}
	Here, and in the following, the subscript $_{\mathrm{tot}}$ indicates the Jones matrix of the complete optical chain, while the subscripts $_{\mathrm{out}}$ and $_{\mathrm{in}}$ refer to the fields that are, respectively, at the output and input of the optical chain.
	
	The result for both polarizations is:
	\begin{equation}\label{eq:Jxy}
		J_{\mathrm{tot},x} = \begin{pmatrix}
			J_{11} & J_{12}\\
			0 & 0\\
		\end{pmatrix}; \quad J_{\mathrm{tot},y} = \begin{pmatrix}
			0 & 0\\
			J_{21} & J_{22}\\
		\end{pmatrix},
	\end{equation}
	
	where the $\theta$ dependence is understood.

	\subsection{Coherency matrix}
	So far, we present the case of a quasi-mono-chromatic fully polarized wave. The CMB signal is only partly polarized and it cannot be easily represented in terms of a Jones vector. The stochastic nature of the quasi-polarized incoming signal requires a statistical description that goes beyond that introduced in the previous section; in other words, the quantity that we can really measure is the time-averaged intensity
	\begin{equation}\label{eq:cohP}
		P \equiv \langle \mathbf{E} \mathbf{E}^{\dagger} \rangle  = \begin{pmatrix} T + Q & U - iV \\ U + iV & T - Q \end{pmatrix},
	\end{equation}
	
	\noindent where $T,Q,U$ are the Stokes parameters that describe the polarization state of the wave. They are defined through the time average of the electromagnetic field:
	\begin{equation} \label{eq:stokes}
		\begin{split}
			T = \langle|E_x|^2\rangle + \langle|E_y|^2\rangle, \quad Q = \langle|E_x|^2\rangle - \langle|E_y|^2\rangle, \quad U = 2 \text{Re}[\langle E_x^{*}E_{y} \rangle], \quad V = 2\text{Im}[\langle E_x^{*}E_{y} \rangle] \,.
		\end{split}
	\end{equation}
	
	So, for the observed polarized signal:\\
	\begin{equation}\label{eq:Pobs}
		\begin{split}
			P_{\text{out}} &= \begin{pmatrix} T + Q & U - iV \\ U + iV & T - Q \end{pmatrix}_{\text{out}} = \langle \mathbf{E_{out}} \mathbf{E_{out}}^{\dagger} \rangle =  \langle J_\mathrm{tot} \mathbf{E_{in}} \mathbf{E_{in}}^{\dagger} J^{\dagger}_\mathrm{tot} \rangle = J_\mathrm{tot} \begin{pmatrix} T + Q & U - iV \\ U + iV & T - Q \end{pmatrix}_\mathrm{in} J^{\dagger}_\mathrm{tot} .
		\end{split}
	\end{equation}
	
	The signal collected by a total power detector is proportional to the Stokes parameter, $T$, which can be obtained as half the trace of $P_\mathrm{out}$. Plugging in Eq.~\ref{eq:Pobs} each of the expressions for $J_{\mathrm{tot},(x/y)}$ given in Eq.~\ref{eq:Jxy} and taking ${[P_\mathrm{out}]}/2$, we obtain the expression for the total power collected by the $x/y$ oriented detectors, $d_{\mathrm{obs},(x/y)}$, as follows:
	\begin{subequations}\label{eq:power}
		\begin{align}
			d_{\mathrm{obs},x}&=\frac{1}{2}\left(\left|J_{11}\right|^2+\left|J_{12}\right|^2\right) T + \frac{1}{2}\left(\left|J_{11}\right|^2-\left|J_{12}\right|^2\right) Q  +  \Re{\left(J_{11}J_{12}^{*}\right) U}  + \Im{\left(J_{11}J^{*}_{12}\right)} V \label{eq:power1},\\
			d_{\mathrm{obs},y}&=\frac{1}{2}\left(\left|J_{21}\right|^2+\left|J_{22}\right|^2\right) T + \frac{1}{2}\left(\left|J_{21}\right|^2-\left|J_{22}\right|^2\right) Q +  \Re{\left(J_{22}J_{21}^{*}\right) U} - \Im{\left(J^{*}_{21}J_{22}\right)} V \label{eq:power2} .
		\end{align}
	\end{subequations}
	In the case of an ideal HWP, Eqs.~\ref{eq:power1}-\ref{eq:power2} become:
	\begin{subequations}\label{eq:power_id}
		\begin{align}
			d_{\mathrm{obs},x}^\mathrm{ideal}&=\frac{1}{2}\left[T+\cos(4\theta)\,Q + \sin(4\theta)\,U\right],\label{eq:power_id1}\\
			d_{\mathrm{obs},y}^\mathrm{ideal}&=\frac{1}{2}\left[T-\cos(4\theta)\,Q - \sin(4\theta)\,U\right].\label{eq:power_id2}
		\end{align}
	\end{subequations}
	
	From Eq.~\ref{eq:power_id}, it is clear that the effect of a rotating HWP is to modulate the detected signal from an input linear polarization four times per rotation of the plate.
	
	A useful decomposition of the coherency matrix that  is subsequently shown to be useful is given in terms of the Pauli matrices:
	\begin{equation}
		P = \langle \mathbf{E} \mathbf{E}^{\dagger} \rangle  = T \sigma_T + Q \sigma_Q + U \sigma_U + V \sigma_V ,
	\end{equation}
	
	\noindent where:
	\begin{equation}
		\sigma_T = \begin{pmatrix} 1 & 0 \\ 0 & 1 \end{pmatrix} ,  \sigma_Q = \begin{pmatrix} 1 & 0 \\ 0 & -1 \end{pmatrix} , \sigma_U = \begin{pmatrix} 0 & 1 \\ 1 & 0 \end{pmatrix} , \sigma_V = \begin{pmatrix} 0 & -i \\ i & 0 \end{pmatrix}. 
	\end{equation}
	
	It is also useful to express the Stokes vector $\mathbf{s} = (T, Q, U, V)$ in terms of the elements of the coherency matrix:
	\begin{equation}
		\mathbf{s}=A\mathbf{P}, \quad A = \begin{pmatrix} 1 & 0 & 0 & 1 \\
			1 & 0 & 0 & -1\\ 
			0 & 1 & 1 & 0\\ 
			0 & \imath & -\imath & 0\\ \end{pmatrix}, 
		\quad \mathbf{P} = \begin{pmatrix} P_{11} \\
			P_{12}\\ 
			P_{21}\\ 
			P_{22}\\ \end{pmatrix},
	\end{equation}
	
	\noindent where $\mathbf{P}=\langle E \times E^\dagger \rangle$ is the Kronecker product of the incoming signal with itself.
	Although the incoming signal is no longer represented as a Jones vector as it was for the fully polarized wave, we note that the Jones formalism still allows a full description of the effects of the train of optical elements on the incoming signal.
	
	\subsection{Mueller matrix formalism} \label{sec:Muell_formalism}
	To express directly how the Stokes parameters, $\mathbf{s} = (T, Q, U, V),$ get transformed by the observation, the Mueller formalism can be adopted. Analogously to the Jones formalism, the observed parameters become $\mathbf{s}_{\text{obs}} = M \mathbf{s}$, where $M$ is the Mueller matrix of the whole optical element.  
	
	In moving on from the Jones to the Mueller matrix, we can easily see that
	\begin{equation}\label{eq:jones2mueller}
		\begin{split}
			&\mathbf{P_{out}} = (J_\mathrm{tot} \times J_\mathrm{tot}^\dagger) \mathbf{P_{in}} \rightarrow (A^{-1} s_\mathrm{out}) = (J_\mathrm{tot} \times J_\mathrm{tot}^\dagger) (A^{-1} s_\mathrm{in})  \\
			&s_\mathrm{out} = A (J_\mathrm{tot} \times J_\mathrm{tot}^\dagger) A^{-1} s_\mathrm{in} = M s_\mathrm{in},\quad \text{where} \quad M\equiv A (J_\mathrm{tot} \times J_\mathrm{tot}^\dagger) A^{-1} 
		\end{split}
		.\end{equation}
	
	In a similar fashion, using the decomposition of the coherency matrix in terms of the Pauli matrices, we can show that: 
	\begin{equation} \label{eq:jones2mueller2}
		M_{ij} = \frac{1}{2} \text{Tr}(\sigma_i J_\mathrm{tot} \sigma_j J_\mathrm{tot}^{\dagger})  \, ,
	\end{equation}
	\noindent where $i,j = \{T,Q,U,V\}$.\\
	It is easy to show that the Mueller matrix of an ideal HWP, for a spinning angle $\theta = 0$ is simply:
	\begin{equation}\label{eq:muellerid}
		M_{HWP}^\mathrm{ideal} = \begin{pmatrix}
			1   & 0 &0 &0\\
			0& 1 & 0 &0\\
			0 & 0 & -1 & 0\\
			0 & 0 & 0 & -1\\
		\end{pmatrix}.
	\end{equation}
	
	In the case of a non-ideal HWP, the elements along the diagonal will deviate from unity, and the off-diagonal elements could be also populated. The most general expression of the Mueller matrix of a realistic HWP is expressed as:
	\begin{equation}\label{eq:mueller}
		M_{HWP} = \begin{pmatrix}
			T_1 & \rho_1 & a_1 & b_1\\
			\rho_2 & T_2 & a_2 & b_2\\
			a_3 & a_4 & c_1 & -s_1\\
			b_3 & b_4 & s_2 & c_2\\
		\end{pmatrix}
		.\end{equation}
	
	By transforming the most general Jones matrix in Eq.~\ref{eq:realistic} according to Eq.~\ref{eq:jones2mueller} or Eq.~\ref{eq:jones2mueller2}, we can find the relation between the Mueller matrix elements and the parameters of the HWP non-idealities introduced in the Jones formalism. The complete expression of the Mueller matrix elements can be found in Appendix~\ref{app:fullM}. Here, we would like to make note of the following: In the same framework as that of Eq.~\ref{eq:muellerid}, if no cross-polarization is present (i.e., $\zeta_{1}=\zeta_2=0), $  then the Mueller matrix in Eq.~\ref{eq:mueller} becomes block-diagonal, with $a_i=b_i=0$ for $i=1,2,3,4$. In addition, we obtain $T_1=T_2\equiv T$, $\rho_1=\rho_2\equiv\rho$, $c_1=c_2\equiv c$, and $s_1=s_2\equiv s$. This is the expression that can be commonly found in the literature \citep[compare e.g.,][]{Bryan2010}. 
	
	Similarly to the Jones formalism, the Mueller matrix for the complete optical system is simply given by the multiplication of the individual matrices 
	\footnote{From the definition of Mueller matrix and the properties of the trace and of the Pauli matrices, it can be shown that the Mueller matrix of the product of Jones matrices is equivalent to the product of the corresponding Mueller matrices:
		\[
		\begin{split}
			&M_1 M_2 = M_{1,ij} M_{2,jk} = \frac12 \text{Tr}(\sigma_i J_1 \sigma_j J_1^\dagger) \frac12 \text{Tr}(\sigma_j J_2 \sigma_k J_2^\dagger) =  \frac12 \text{Tr}( J_1^\dagger \sigma_i J_1 \sigma_j) \frac12 \text{Tr}(\sigma_j J_2 \sigma_k J_2^\dagger) = \frac14 (J_1^{\dagger da} \sigma^{ab}_i J^{bc}_1 \sigma^{cd}_j )  (\sigma^{ef}_j J^{fg}_2 \sigma^{gh}_k J_2^{\dagger he}) =\\
			&= \Bigg(\sum_j \sigma^{cd}_j \sigma^{ef}_j = 2 \delta^{cf} \delta^{de} \Bigg) = \frac12 (J_1^{\dagger da} \sigma^{ab}_i J^{bc}_1 J^{cg}_2 \sigma^{gh}_k J_2^{\dagger hd})  = \frac12 \text{Tr}(\sigma_i J_1 J_2 \sigma_j (J_1 J_2)^\dagger) = M_{12} \, .
		\end{split}
		\]      
		\noindent       
		So, $M_{pol}M_{J_{rot}^{T}}M_{J_{HWP}}M_{J_{rot}} = M$ as defined above.}. 
	
	If we set the matrix elements of such a Mueller matrix to:
	\[
	M_{x/y} = \begin{pmatrix}
		M^{TT}  & M^{TQ} & M^{TU} & M^{TV}\\
		M^{QT}& M^{QQ} & M^{QU} & M^{QV}\\
		M^{UT} & M^{UQ} & M^{UU} & M^{UV}\\
		M^{VT} & M^{VQ} & M^{VU} & M^{VV}\\
	\end{pmatrix},
	\]
	
	\noindent where the subscript $x/y$ implies that the optical train ends with a polarizer along $x/y$, and the same subscript is understood in each of the matrix elements. 
	
	The total power collected by a single detector -- being proportional to the Stokes parameter T -- corresponds to taking the first row of the Mueller matrix and multiplying it by the Stokes vector of the input signal. The general expression of the signal obtained by one detector is:
	\begin{equation}\label{eq:todM}
		d_{\mathrm{obs},(x/y)} = M^{TT}_{x/y} \,T + M^{TQ}_{x/y} \,Q + M^{TU}_{x/y} \,U + M^{TV}_{x/y} \,V.
	\end{equation}
	
	From Eq.~\ref{eq:jones2mueller},  we can see the equivalence between each of the coefficients $M^{TT}_{x/y}, M^{TQ}_{x/y}, M^{TU}_{x/y}, M^{TV}_{x/y}$ and the terms appearing in Eqs.~\ref{eq:power1}-\ref{eq:power2}, derived from the Jones formalism.
	
	\subsection{Study in context}
	In this work, we only consider the specific case-study of a detector at the boresight collecting signal coming from light hitting the HWP perpendicularly. In the absence of beam convolution, the light rays convolved by the optical system on the detector at boresight are the orthogonal ones \citep{lamagna2021optical}. That is why our Mueller matrix elements depend only on the spinning angle of the HWP and does not depend on the incidence angle, which would be the case in general. To take into account non-orthogonal incidence, we should include the dependence on the angle of incidence in the computation of the Mueller matrix elements \citep{Salatino:2018voz,Essinger_Hileman_2016} and convolve the total matrix by the beam \citep{Duivenvoorden:2020xzm}. In this work, we have chosen a simplified approach neglecting the coupling between beam convolution and HWP non-idealities. This coupling can be a source of additional systematic effects, such as temperature-to-polarization leakage, which are not included in this study and instead deferred to a future work (Patanchon et al., in prep.).
	We refer to \citep[Patanchon et al., in prep.;][]{Salatino:2018voz,DAlessandro:2019snm,Duivenvoorden:2020xzm, Essinger_Hileman_2016}
	for studies that include the effect of slant incidence.
	
	By considering orthogonal incidence only, we are implicitly assuming a symmetric beam. However, we note that even if the beam is asymmetric in its shape, it is symmetrized to some extent due to the scanning strategy. This symmetrization effect further motivates our choice of restricting the study to orthogonal incidence.
	
	Based on Eq.~\ref{eq:todM} combined with Eq.~\ref{eq:mueller}, we may notice that in the ideal case, we have $M^{TV} = 0$. An optical system employing a realistic (non-ideal) HWP allows detection of V-mode signal~\citep{Nagy:2017csq}. 
	However, in this work, we will restrict to the case $V=0$ as expected in the standard cosmological model. We note that several mechanisms have been proposed to generate a certain amount of CMB circular polarization~\citep{Lembo:2020ufn,Zarei_2010,Alexander_2009,Alexander_2020,Sadegh:2017rnr,PhysRevD.99.043501,Vahedi_2019,PhysRevD.100.043516}. Nevertheless, the predicted signal is very faint, and therefore justifies our choice of assuming $V=0$ in the next section.
	
	A final note before we move on to discuss other aspects of our analysis. In principle,
	for the simple case we are studying here (i.e., a non-depolarizing optical system or a normal incidence) we could have worked with the Jones formalism, provided that the input and output signals were described in terms of the coherency matrix. Nevertheless, we decided to switch to the Mueller formalism since it provides a more direct handle to the Stokes parameters. 
	
	\section{Application to future CMB missions: Scanning strategy and mapmaking}\label{sec:strategy}
\begin{figure}[!htbp] 
		\centering
		\includegraphics[width=0.5\textwidth]{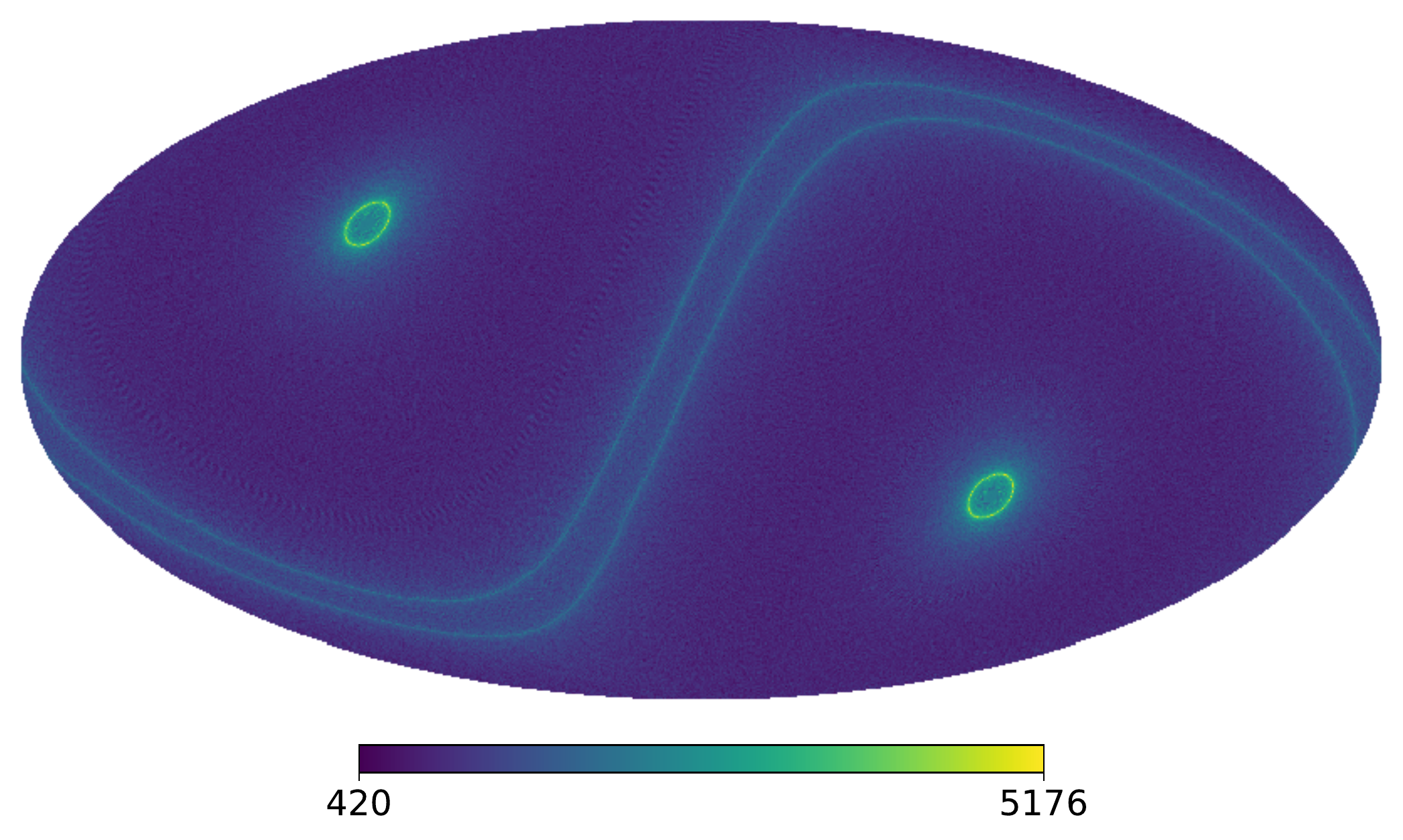}
		\caption{Map displaying the number of samples collected in each pixel, for a pair detectors at boresight. {\ttfamily Healpix} resolution: $N_{side}$ = 512.}\label{fig:hitmap}
	\end{figure}

	In the previous section, we laid down the mathematical formalism to describe the effects of a continuously rotating, non-ideal HWP. In this section, we present the experimental setup we want to investigate, in which such a HWP is employed. We are interested in quantifying the impact of HWP non-idealities in the context of future CMB observations. In particular, we focus on simulating the performance of a HWP on board of a LiteBIRD-like satellite experiment. 
	LiteBIRD \citep{Hazumi:2021yqq} is a satellite mission expected to be launched in the late 2020s, whose main scientific target is the detection of an inflationary signal with a precision on $r$ of $\sigma_r \lesssim 10^{-3}$. LiteBIRD will perform a full-sky survey over three years at the Sun-Earth Lagrangian point L2, using three telescopes (Low-Frequency Telescope LFT, Medium-Frequency Telescope MFT, and High-Frequency Telescope HFT) that observe in 15 frequency bands between 34 and 448 GHz . Each telescope will use a HWP as a polarization modulator. The parameters defining the LiteBIRD scanning strategy are listed in Table \ref{tab:litebird} and a map of the expected satellite footprint is shown in Fig. \ref{fig:hitmap}.
	
	\begin{table}
		\begin{center} 
			\caption{Parameters defining the LiteBIRD scanning strategy}    \label{tab:litebird}
			\begin{tabular}{c|c}
				\hline
				Sampling rate (Hz) & $19$ \\
				Mission time (months) & $ 36 $ \\
				
				HWP spin velocity (Hz) LFT/MFT/HFT& $ 0.77/0.65 /1.02$ \\
				Precession angle ($^\circ$)& $45 $ \\
				Boresight angle ($^\circ$)& $50  $\\
				Precession velocity (rad/min) & $0.033$ \\  
				Satellite spin (rad/min) & $0.31$\\  
				\hline          
			\end{tabular}
			
		\end{center}
		{\raggedright \textbf{Notes.} When we limit to the single-frequency study in this work, we set the HWP spin velocity to the MFT value   (LiteBIRD collaboration, (in prep.).  \par}
	\end{table}
	
	We developed a software package that simulates a realistic scanning strategy of a satellite mission, and subsequently reconstructs maps of the $T$, $Q,$ and $U$ Stokes parameters from the simulated observations.   As noted earlier in this paper, we considered a single pair of polarization-sensitive detectors located at boresight (i.e., perfectly centered on the instrument focal plane). Both detectors share a view of the same sky patch, but they are oriented at $90^\circ$ with respect to each other, so as to remain sensitive to orthogonal polarization directions. The full optical chain as viewed by the sky signal entering the telescope is then composed by a continuously rotating HWP followed by a pair of orthogonal polarization sensitive detectors. We also account for the relative orientation of the telescope with respect to the local coordinate system that locally identifies $Q$ and $U$. 
	The Mueller matrix of the full optical chain is obtained as the product of the individual Mueller matrices:
	\begin{equation}\label{eq:mueller_optsys}
		M_\mathrm{full,j}(t)= M_{pol,j}M_\mathrm{rot}^{T}(t)M_\mathrm{HWP}M_\mathrm{rot}(t)M_\psi(t),\,j=x,y,
	\end{equation}
	
	where the time dependence has been made explicit where relevant. As explained in the previous section, $M_{pol,j}$ is the Mueller matrix for the polarizers along the $j = x,y$ direction, $M_\mathrm{rot}$ is the Mueller version of $R(\theta)$,
	$M_\mathrm{HWP}$ is the Mueller matrix of a non-ideal HWP (defined as in Eq. \ref{eq:mueller}). Finally, $M_\psi$ is the Mueller version of the rotation matrix $R(\psi)$, that takes into account the angle $\psi$ between the instrument and the local coordinate system on the sky.
	The total power collected by a single detector at a given time, $t$, also known as time-ordered data (TOD), is then
	
	\begin{equation}
		d_\mathrm{obs}(t)=\sum_j M^{TT}_\mathrm{full,j} \,T + M^{TQ}_\mathrm{full,j} \,Q + M^{TU}_\mathrm{full,j} \,U.
	\end{equation}
	
	Since the instrument has finite angular resolution, the sky is discretized in small patches (pixels). At a given time, $t_i$, one pixel, $p,$
	is observed. We follow the \texttt{HEALPix} pixelization scheme \citep{Gorski:2004by}. The sky is divided into $12\times N_\mathrm{side}^2$ pixels, where $N_\mathrm{side}$= 512 is chosen in such a way that the size of each pixel is smaller than the angular resolution of the experiment \citep[0.5 $\deg$ ~at 100 GHz;][]{Hazumi:2021yqq}.
	 Ignoring beam convolution\footnote{Our treatment assumes that the Mueller matrix elements are not affected by beam convolution and we can safely convolve input maps with Gaussian beams prior to the simulated observation.}, the total power collected by a single detector at a given time $t_i$ is then
	\begin{equation}\label{eq:dobsti}
		\begin{split}
			&d_\mathrm{obs}(t_i)= M^{TT}_{\mathrm{full},p}(t_i) \,T(p) + M^{TQ}_{\mathrm{full},p}(t_i) \,Q(p) + M^{TU}_{\mathrm{full},p}(t_i) \,U(p)+n_i,
		\end{split}
		\end{equation}
	
	where the sum of $j = x,y$ is now understood. We also allow for the possibility of instrumental noise, $n_i$, to be added to the $i$-th time sample. We note that the Mueller matrix of the optical system also depends on the observed pixel through the relative orientation with respect to local coordinate system. The TOD equation can be arranged in a matrix notation:
	\begin{equation} \label{eq:todint}
\begin{split}
	&\mathbf{d}_\mathrm{obs}(t)  = \begin{pmatrix} \vdots \\ d_\mathrm{obs}(t_i) \\ \vdots \end{pmatrix} = \begin{pmatrix}  \ddots & \vdots & \vdots & \vdots & \vdots & \vdots & \udots \\ ... & 0 &  M^{TT}_{\mathrm{full},p}(t_i) & M^{TQ}_{\mathrm{full},p}(t_i) & M^{TU}_{\mathrm{full},p}(t_i) &  0 &... \\ 
		\udots & \vdots & \vdots & \vdots & \vdots & \vdots & \ddots  \end{pmatrix} \begin{pmatrix} \vdots \\ T(p) \\ Q(p)\\ U(p) \\  \vdots \end{pmatrix} + \begin{pmatrix} \vdots \\ n_i \\ \vdots \end{pmatrix}
	\equiv  \mathbf{A}(t) \mathbf{m}_\mathrm{in} + \mathbf{n}(t),
\end{split}
		\end{equation}
	
	\noindent where $A$ is the "pointing matrix" with the following dimension: 
	N$_{\rm samples}$  $\times$ (3 $\times$ N$_{\rm pixels}$ ), $\mathbf{m}_\mathrm{in}$ is the vector of Stokes parameters with the dimension: (3 $\times$ N$_{\rm pixels}$), and $\mathbf{n}$ is the vector of instrumental noise contributions with dimension (N$_{\rm samples}$). The rows of the pointing matrix have non-zero elements only for the samples $i$ in which the pixel $p$
	is observed. Finally, $\mathbf{d}_\mathrm{obs}(t)$ is the full TOD vector, with the dimension (N$_{\rm samples}$). The number of samples is easily computed from the mission duration and the data sampling rate (see Table~\ref{tab:litebird}). Equation \ref{eq:todint} can be inverted to reconstruct the Stokes maps from the TOD with the mapmaking procedure, as, for example, in \citet{Tegmark:1996qs}, \citet{Natoli:2001tb}, \citet{Keihanen:2004yj}. In the case of non-correlated noise (i.e., $\langle n n^T \rangle = \sigma^{2} \mathbb{I}$, where $\sigma$ is the uniform noise standard deviation), it can be shown that the reconstructed sky signal in matrix form is 
	\begin{equation} \label{eq:map-making}
		\mathbf{m}_\mathrm{out} = (\mathbf{B}^{T}\mathbf{B})^{-1}\mathbf{B}^{T} \mathbf{d}_\mathrm{obs} = (\mathbf{B}^{T}\mathbf{B})^{-1}\mathbf{B}^{T} \mathbf{A} \, \mathbf{m}_\mathrm{in} + (\mathbf{B}^{T}\mathbf{B})^{-1}\mathbf{B}^{T} \mathbf{n},
	\end{equation}
	where $\mathbf{B}$ is the estimated pointing matrix, usually constructed from the actual attitude of the telescope and from a pre-launch measurement of the instrument optical elements.

	In the next sections, we will use the following formalism: $A$ is the ``real'' pointing matrix and includes all the HWP systematic effects that might affect our data (we will also refer to it as ``TOD HWP''); $B$ is the estimated pointing matrix, which we refer to as the ``solver'' matrix or as ``map-making HWP." Here, $B$ is to be used in the mapmaking process and we go on to construct it either as identical to $A$ (to correctly recover the input sky signal at the right hand side of Eq.~\ref{eq:map-making}) or to be different from $A$ (to propagate the effect of unaccounted systematics). 
	
	Armed with this basic formalism,
	we are now ready to follow the steps implemented in the code:

	       \begin{enumerate}
	                \item at each time step $t_i$, the observed pixel $p$
	                is identified, the signal $d_\mathrm{obs}(t_i)$
	                as given in Eq. \ref{eq:dobsti} is computed (using the matrix $A$) for both detectors observing that pixel;
	                \item from Eq. \ref{eq:map-making}, the two quantities $(B^{T}B)_{p}$ and $(B^{T} d_\mathrm{obs})_{p}$ 
	                are computed for the pixel $p$.
	                It should be noted that the algorithm does not require storage of the TOD vector;
	                \item every time a pixel $p$
	                is observed, the two quantities above are summed to those already computed in previous steps for the same pixel. The number of samples falling in each pixel is also stored to produce a coverage map (Fig.~\ref{fig:hitmap}) at the end;
	                \item we cycle over the first three points for all the time samples collected by the instruments;
	                \item at the end of the mission time, the Stokes maps are estimated using Eq. \ref{eq:map-making}.
	        \end{enumerate}
	
	In the next few sections, we demonstrate how we applied the algorithm above to two case studies to quantify the effects of HWP systematics in the context of future satellite missions. In Sect.~\ref{sec:mono}, we consider the simple case of single-frequency observations of a CMB-only sky with white noise.
	This allows us to easily understand the impact of each class of HWP non-idealities on the output signal. In Sect.~\ref{sec:cromo}, we consider a more realistic scenario that consists in multi-frequency observations of a more complex sky with the CMB signal contaminated by the presence of (frequency-dependent) foreground emission. Since the systematic effects treated in this paper do not cause temperature-to-polarization leakage, we only focus on polarized emission. The analysis setup and results for both the mono-chromatic and multi-frequency studies are described in detail in the following sections. 
	
	\section{Single-frequency case}\label{sec:mono}
	In this section, we describe how we applied the algorithm described in Sect.~\ref{sec:strategy} to a simple case study. We simulated observations with a future CMB satellite. We assumed perfectly monochromatic detectors, that is, we observed the sky at one frequency. The input sky is given by the CMB signal only, both in temperature and polarization. We also assumed a simple model for the instrumental noise, properly rescaled to take into account that our simulated observations only employ two detectors. Finally, we assumed that no systematics but those related to HWP non-idealities are present. To quantify the impact of HWP non-idealities, we compared the
	$BB$ power spectrum residuals 
	to the ideal $BB$ power spectrum that would be observed in the absence of HWP systematics. 
	
	\subsection{Input sky and experimental setup}\label{sec:mono_setup}
	The input sky is composed of the CMB signal only. We computed $100$ $T,\,Q,\,U$ map realizations from the same fiducial set of $TT,\,TE,\,EE,\,BB$ CMB power spectra. The latter were computed with the default values of the Boltzmann solver \texttt{CAMB}~\citep{Lewis:1999bs}. We assumed a flat $\Lambda$CDM cosmology with three families of active neutrinos with total mass 0.06 eV. We note, however, that the choice of a different cosmology would have had negligible impact on the results presented in this section. We allowed for a non-zero value of the tensor-to-scalar ratio, which we set to $r=0.003$. We use the lensed version of the spectra as generated by \texttt{CAMB}, to account for the extra variance from lensing particularly relevant for the $BB$ signal. Maps were generated with the \texttt{HEALPix} routine \texttt{synfast} as implemented in the python package \texttt{healpy}. The generation of a large number of CMB realizations is needed to account for cosmic variance: in the absence of systematic effects, the ensemble average of the observations is expected to reproduce the fiducial input spectra \citep{Gerbino:2019okg}.
	
	We adopted the publicly available instrumental specifications of the future LiteBIRD satellite (summarized in Table~\ref{tab:litebird}). We set $N_\mathrm{side}=512$ and smooth the signal with a $FWHM=30.8\,\mathrm{arcmin}$ gaussian beam to simulate a LiteBIRD-like angular resolution at $150\,\mathrm{GHz}$ \citep[Table \ref{tab:bands};][]{Hazumi:2021yqq}.
	When generating maps, we also take into account the effect of the pixel window function. Each of the 100 sky realizations is used as the input sky for the scanning and mapmaking algorithm described in the previous section. We include experimental noise to highlight differences in the noise bias due to different choices of the map-making matrix $B$. Since the main focus of this work is to study the effects of HWP non-idealities, we only consider a simple noise contribution that is isotropic 
	and uncorrelated.
	These properties translate to a white noise spectrum in harmonic space: $N_\ell=\Sigma_m \sigma^2/(2\ell+1)=\sigma^2$. In practice, $n_i$ in Eq. \ref{eq:dobsti} is drawn from a zero-mean Gaussian distribution with variance $\tilde{\sigma}^2$. The noise variance is determined from the intrinsic detector sensitivity $NET$ and the number of samples $t_\mathrm{p}$ in which each pixel is observed.
	Assuming the specifications in Table~\ref{tab:bands}, we obtain the following:  $\tilde{\sigma}= 3.16 \,\mathrm{\mu K}$.
	
	Each sky realization was observed twice. First, we ran the code setting the pointing matrix $A$ to be equal to the solver matrix $B$, both with an ideal HWP. In a second run, we instead imposed $B$ to be different from $A$; thus, $A$ takes into account a non-ideal HWP while $B$ can either consider an ideal or a non-ideal HWP. The exact values of the HWP parameters entering $A$ and $B$ are given in Sect.~\ref{subsec:mono_res}. In both cases, the output of the code consists of a reconstructed map that is the sum of the observed CMB signal and instrumental noise.
	In total, we have two sets of 100 output maps: one set of 100 ideal output maps and another set of 100 realistic output maps. We computed the output $TT,\,TE,\,EE,\,BB$ spectra from each output map employing the \texttt{HEALPix} routine \texttt{anafast} from the \texttt{healpy} python package. When generating the output spectra, we assume full-sky observations and we correct for the beam smoothing effect and for the pixel window function. As stated at the beginning of this section, we focus on $BB$ residuals to quantify the impact of HWP non-idealities. Therefore, going forward, we only focus on $BB$ spectra $C_\ell^\mathrm{BB}$. In total, we have two sets of 100 output spectra: a set of $i=1,2,...,100$ ideal spectra $C_{\ell,i}^\mathrm{BB}$(ideal), and another set of $i=1,2,...,100$ realistic spectra $C_{\ell,i}^\mathrm{BB}$(realistic) $= C_{\ell,i}^\mathrm{BB}$(w/ systematics)$+N^{BB}_{\ell,i}$ \footnote{The output maps produced by these simulation are affected by noise, so their power spectrum $C_{\ell,i}$(realistic) can be written as the sum of the CMB power spectrum affected by the systematics and the white noise power spectrum, neglecting any chance-correlation.}. A schematic picture of the procedure described above is depicted in Fig.~\ref{fig:code_scheme}.
	
	\tikzstyle{startstop} = [rectangle, rounded corners, minimum width=3cm, minimum height=1cm,text centered, text width=3cm, draw=black, fill=red!30]
	\tikzstyle{process} = [rectangle, minimum width=3cm, minimum height=1cm, text centered, draw=black, fill=orange!30]
	\tikzstyle{arrow} = [thick,->,>=stealth]
	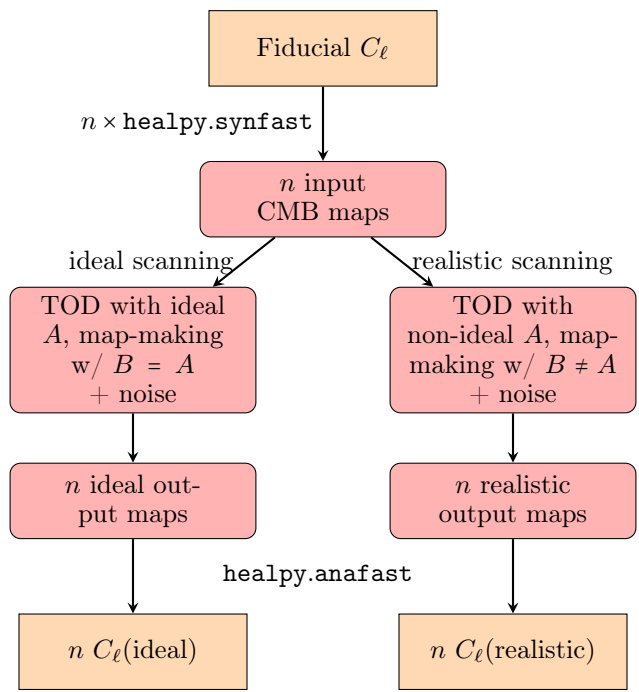
\begin{figure}[!htbp] 
		\centering
		\begin{tikzpicture}[node distance= 2 cm] 
			\node(start)[process]{Fiducial $C_{\ell}$};
			\node (pro0) [startstop, below of=start] {$n$ input CMB maps};
			\node (pro1) [startstop, below of=pro0, xshift=-2.5 cm] {TOD with ideal $A$, map-making w/ $B=A$ \\+ noise};
			\node (pro2) [startstop, below of=pro0, xshift=+2.5 cm] {TOD with non-ideal $A$, map-making w/ $B \neq A$ \\+ noise };
			\node (pro3) [startstop, below of=pro1] {$n$ ideal output maps};
			\node (pro4) [startstop, below of=pro2] {$n$ realistic output maps};
			\node (pro5) [process, below of=pro3] {$n$ $C_{\ell}$(ideal)};
			\node (pro6) [process, below of=pro4] {$n$ $C_{\ell}$(realistic)};
			\draw [arrow] (start) -- node[anchor=east] {$n \times \mathtt{healpy.synfast}$} (pro0);
			\draw [arrow] (pro0) -- node[anchor=east] {ideal scanning} (pro1);
			\draw [arrow] (pro0) -- node[anchor=west] {realistic scanning} (pro2);
			\draw [arrow] (pro1) -- node[anchor=east] {} (pro3);
			\draw [arrow] (pro2) -- node[anchor=west] {} (pro4);
			\draw [arrow] (pro3) -- node[anchor=west] {$\qquad \quad \mathtt{healpy.anafast}$} (pro5);
			\draw [arrow] (pro4) -- node[anchor=east] {} (pro6);
		\end{tikzpicture}
		\caption{Scheme of the procedure for the monochromatic analysis. From a set of $n=100$ input maps, we obtain two sets of ideal and realistic output spectra, depending on whether we allow for the TOD matrix $A$ to be equal to or different from the mapmaking matrix $B$, respectively.}
		\label{fig:code_scheme}
	\end{figure}

	To get the $BB$ residuals due to systematics, we first need to noise-debias the observed spectra. The noise bias is obtained as the average over noise spectra computed from 100 noise maps drawn from the noise covariance matrix $\mathcal{N}\equiv \sigma^2 \left(B^T B\right)^{-1}$.
	Clearly, different choices of the $B$ matrix would lead to different noise on the maps.
	
	We considered two different scenarios. In both scenarios, the map-making matrix $B$ is kept fixed while we consider different choices for the pointing matrix $A$, so to illustrate the effects of unaccounted systematics. 
	
	First, we took $B$ to be the map-making matrix of an ideal optical system, that is, the solver as computed from Eq. \ref{eq:mueller_optsys} when using Eq. \ref{eq:muellerid} for the ideal HWP. We reconstructed the output map for different choices of the pointing matrix $A=A(h_1,h_2,\zeta_1,\zeta_2,\beta, \chi_1,\chi_2)$ to highlight the impact of each class of HWP non-idealities represented by the parameters $h_1,h_2,\zeta_1,\zeta_2,\beta, \chi_1,\chi_2$. In detail, we considered the following three classes: non-ideal transmittance with $A\equiv A(h_1,h_2=\mathrm{const},\zeta_1=\zeta_2=0,\beta=0, \chi_1=\chi_2 =  0)$, non-vanishing cross-polarization with $A\equiv A(h_1=h_2=0,\zeta_1,\zeta_2=\mathrm{const},\beta=0, \chi_1,\chi_2 =\mathrm{const})$, and non-ideal phase-shift with $A\equiv A(h_1=h_2=0,\zeta_1=\zeta_2=0,\beta=\mathrm{const}, \chi_1=\chi_2 =  0)$. We extended the second class by perturbing also the phases $\chi_{1,2}$ of the cross-polarization terms, even though in general they can be reabsorbed in a redefinition of $\zeta$. Within each class, we explored different values of the non-vanishing non-ideal parameters. A summary of these values is reported in Table~\ref{tab:monoA}.
	
	\begin{table*}
		\begin{center}
			\caption{Parameter values adopted to build the pointing matrix $A$ in the case of an ideal solver matrix $B$}\label{tab:monoA}
			\begin{tabular}{c|c c c|c c c|c c c|c c}
				& \multicolumn{3}{c|}{$h_1,h_2 \neq 0$}  &\multicolumn{3}{c|}{$\zeta_1,\zeta_2 \neq 0$} &\multicolumn{3}{c|}{$\beta \neq 0$}&\multicolumn{2}{c}{$\chi_1,\chi_2 \neq 0$}\\
				\hline
				$h_1$ &-0.1&-0.15&-0.05&$0$&0&0  &$0$&0&0&0&0\\
				$h_2$ &-0.1&-0.05&-0.25 &$0$&0&0&$0$&0&0&0&0\\
				$\zeta_1$ &0&0&0  &0.1&0.15&0.18&$0$&0&0&0.1&0.1\\
				$\zeta_2$ &0&0&0  &0.1&0.05&0.15&$0$&0&0&0.1&0.1\\
				$\beta$ &0&0&0  &0&0&0  &10$^\circ$& 20$^\circ$&30$^\circ$&0&0\\
				$\chi_1$ &$-$&$-$&$-$  & 0&0&0  &$-$&$-$&$-$&10$^\circ$&-30$^\circ$\\
				$\chi_2$ &$-$&$-$&$-$  &0&0&0  &$-$&$-$&$-$&-20$^\circ$&20$^\circ$\\
			\end{tabular}
		\end{center}
		{\raggedright \textbf{Notes.} Each column specifies the combination of parameters used in each case. The sign $-$ refers to the fact that for null $\zeta$ their phases are undefined. \par}  
	\end{table*}
	
	A second setup was then considered. We took $B$ to be the solver matrix of a more realistic optical system, that is, we allowed for the non-ideal parameters to be non-vanishing one at a time. We again computed the output maps for different choices of the pointing matrix $A$, similarly to what was done in the previous case.  Values employed to build the pointing matrix are reported in Table~\ref{tab:monoA2}, where we highlight in boldface the values used to build the map-making matrix $B$ (we use the subscript $s$ for them to indicate the "solver"). The first setup allows for a characterization of the residuals when the HWP systematics are not accounted for in the map-making; the second one is when they are accounted for, but with a mismatch between their estimate in the solver and their actual value in the pointing matrix. 
	In this second setup, we did not consider perturbations of the phases $\chi_{1,2}$,
	as the residuals are mainly driven by the value of $\zeta$.
	
	We adopted exaggerated values for the non-ideal parameters in order to make their effect on the power spectra well visible. In Sect.~\ref{sec:cromo}, we consider more realistic values in order to propagate their effects to $r$.

	\begin{table}
		\begin{center}
			\caption{Parameter values adopted to build the pointing matrix $A$ for the case of a non-ideal solver matrix $B$}\label{tab:monoA2}
			\begin{tabular}{c|c c c|c c c|c c c}
				& \multicolumn{3}{c|}{$h_1,h_2 \neq 0$}  &\multicolumn{3}{c|}{$\zeta_1,\zeta_2 \neq 0$} &\multicolumn{3}{c}{$\beta \neq 0$}\\
				\hline
				$h_1$ &-0.15&$\textbf{-0.1}$&-0.05 &$0$&$0$&$0$  &$0$&$0$&$0$\\
				$h_2$ &-0.05&$\textbf{-0.1}$&-0.02 &$0$&$0$&$0$  &$0$&$0$&$0$\\
				$\zeta_1$ &$0$&$0$&$0$  &0.15&$\textbf{0.1}$&0.2  &$0$&$0$&$0$\\
				$\zeta_2$ &$0$&$0$&$0$  &0.05&$\textbf{0.1}$&-0.05  &$0$&$0$&$0$\\
				$\beta$ &$0$&$0$&$0$  &$0$&$0$&$0$  &$5^\circ$&$\textbf{15}^\circ$&$30^\circ$\\
			\end{tabular}
		\end{center}
		{\raggedright \textbf{Notes.} In boldface, values of the non-ideal parameters used to build $B$ for the case under consideration. The solver matrix $B$ is kept fixed in this single-frequency study. In this setup, we are not considering perturbations of $\chi$. Each column specifies the combination of parameters used in each case. \par}
	\end{table}
	
	\subsection{Results of the single-frequency analysis}\label{subsec:mono_res}
	Here, we present and discuss the results obtained for the single-frequency analysis. The $BB$ power spectra for the two cases of ideal mapmaking matrix $B$ and non-ideal $B$ discussed in the previous section are presented in Figures~\ref{fig:plot_id_B} and~\ref{fig:plot_non_id_B}.  \\
	
	\subsubsection{Ideal map-making matrix $B$}
\begin{figure}[!htbp] 
		\centering
		\includegraphics[width=0.55\textwidth]{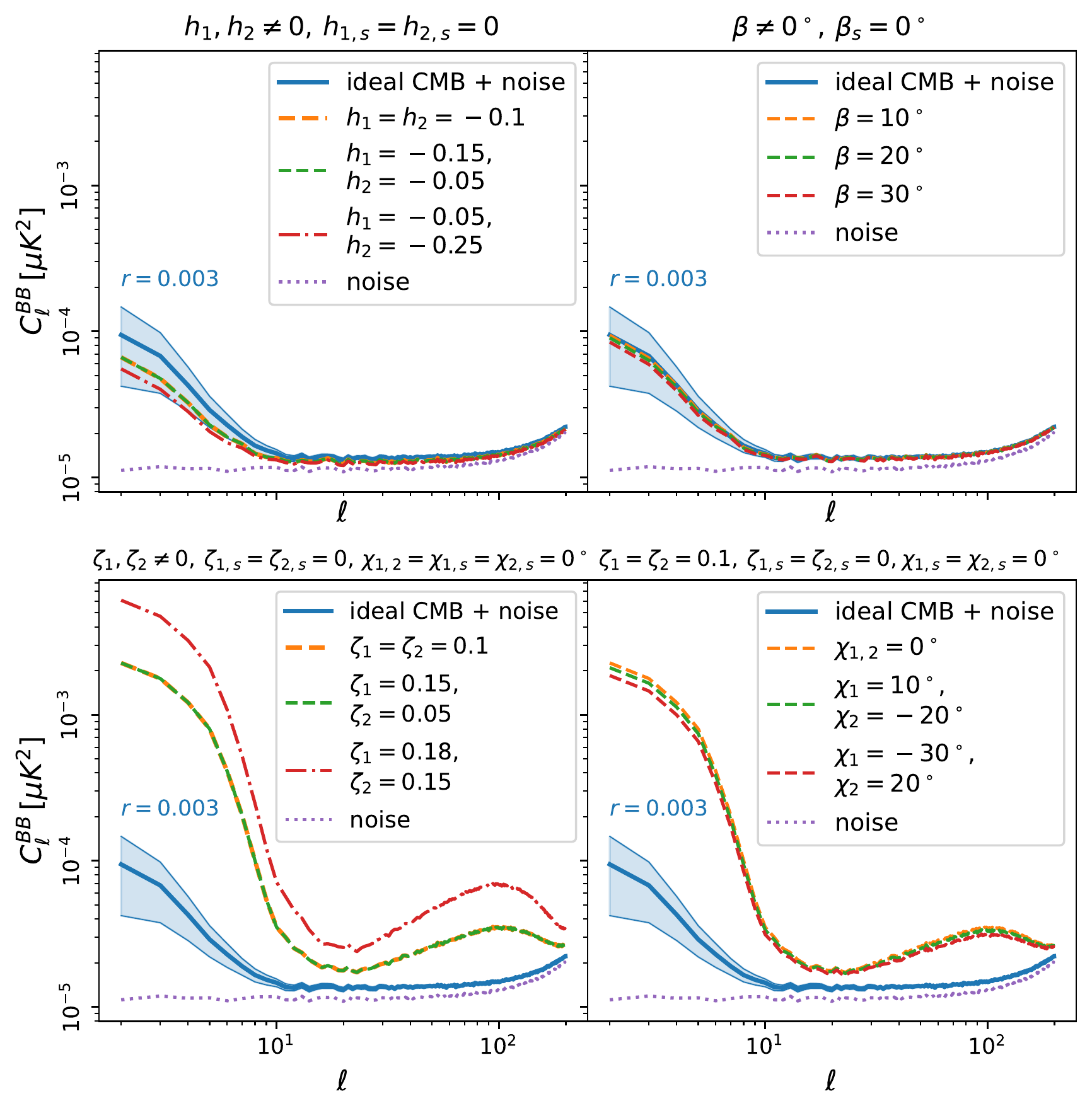}
		\caption{$C^{BB}_{\ell}$ power spectra with CMB only (no foregrounds) and noise simulations. This figure summarizes the results in the case of ideal map-making matrix $B$ (ideal HWP). We report in blue the CMB (with $r = 0.003$) + noise spectra in the ideal case (TOD matrix equal to the mapmaking matrix, $A = B$). The standard deviation of the 100 ideal CMB realizations is shown as a shaded blue region. Spectra from the output maps obtained with the choice $A \neq B \rightarrow$ (systematic+noise) are shown in dashed-dotted and dashed lines. In the case of $h$ (or $\zeta$), the dashed lines have the same $h_1+h_2$ (or $\zeta_1+\zeta_2$), whereas this sum is different in the case shown in the dashed-dotted lines: we see that the dashed lines are overlapping, as the $2 \theta$ terms in Eq.~\ref{eq:Muellerx_h}, \ref{eq:Muellerx_z} are 
			canceled out in the mapmaking process. The noise bias is shown in dotted. In the case of ideal $B$, the noise bias is the same regardless from the parameter perturbed.
			We refer to the main text for a more detailed discussion.}\label{fig:plot_id_B}
	\end{figure}

	The main findings in the case of an ideal solver matrix B are summarised in Fig.~\ref{fig:plot_id_B}, where we show:
	in blue solid line, the average spectrum obtained from 100 realizations of ideal CMB maps, to which we sum the noise bias $ \langle C_{\ell}^\mathrm{BB}$(ideal)$\rangle + \langle N^{BB}_{\ell} \rangle$ (this is our reference spectrum); in colored dashed or dashed-dotted line, we show the average spectrum over 100 CMB realizations affected by noise and one kind of systematics at a time $ \langle C_{\ell}^\mathrm{BB}$(realistic)$\rangle$; in purple dotted line, we show the noise spectrum $\langle N^{BB}_{\ell} \rangle$.
	
	Since the map-making matrix $B$ is fixed to the ideal case, the systematic effects due to non-ideal parameters are not taken into account in the map-making stage. Because of this, the noise spectrum shown in Fig.~\ref{fig:plot_id_B} is the same for all the panels (see Sect.~\ref{sec:mono_setup}). 
	
	Next, we discuss the impact of each class of non-idealities. The parameters $h_1,h_2$ have the effect of shifting the spectra to lower amplitudes. This can clearly be seen by expanding $M^{TX}_x$ (Eq.~\ref{eq:Muellerx}) for small $h$, with all the parameters but $h$ set to zero:
	
	\begin{equation} \label{eq:Muellerx_h}
		\begin{split}
			M^{TT}_{x} &\simeq  \frac12 (1+h_1+h_2)+ \frac12 (h_1-h_2)\cos(2\theta), \\ 
			M^{TQ}_{x} & \simeq  \frac12 (h_1-h_2)\cos(2\theta) + \frac12 \left(1+h_1+h_2 \right) \cos(4\theta), \\
			M^{TU}_{x} & \simeq \frac12 (h_1-h_2)\sin(2\theta) + \frac12\left(1+h_1+h_2 \right) \sin(4\theta). \\
		\end{split}
	\end{equation}
	The top left panel of Fig.~\ref{fig:plot_id_B} shows the effects of $h_{1,2}$ on the $BB$ spectrum. 
	In the dashed lines, we report the results for two combinations of $h_{1,2}$ which share the same value of $h_1 + h_2 = -0.2$ but different $h_1-h_2$. The two dashed curves overlap almost perfectly. This can be explained with the fact that the $2 \theta$ terms in Eq.~\ref{eq:Muellerx_h}, scaled by $h_1-h_2$, are canceled out in the map-making procedure by considering orthogonally polarized detectors\footnote{This still holds also when we consider non-ideal parameters in the solver provided that $h_{1,s} = h_{2,s}$ and $\zeta_{1,s} = \zeta_{2,s}$. Instead, introducing an unbalance between the two axis in the solver let the $2 \theta$ harmonics survive.}.
	
	The case of $\beta \neq 0$ is shown in the top right panel of Fig.~\ref{fig:plot_id_B}. 
	Expanding Eq.~\ref{eq:Muellerx} with respect to $\beta$ and setting the other parameters to zero, we obtain:
	\begin{equation} \label{eq:Muellerx_b}
		\begin{split}
			M^{TT}_{x} & = \frac12,   \\ 
			M^{TQ}_{x} & = \frac14 \left(1-\cos\beta\right) + \frac14  \left(1+\cos\beta\right)\cos(4\theta),\\ 
			M^{TU}_{x} & =  \frac14 \left(1+\cos\beta\right)\sin(4\theta).
		\end{split}
	\end{equation}
	The $4 \theta$ terms are now scaled by $(1+\cos \beta)$, which acts to reduce the output signal when we fix $\beta_s = 0$. Indeed, in the ideal case of $\beta=0$, we should measure a signal with an ideal phase-shift of exactly 180$^\circ$. However, the actual signal is detected with a slightly different phase-shift and a fraction of the input power is not transfered. This can be observed in the plot, where the shift toward lower amplitude of the $BB$ spectrum is more enhanced for higher values of $\beta$ (always smaller than $90^\circ$).
	
	The cases with $\zeta \neq 0$ are shown in the lower panels of Fig.~\ref{fig:plot_id_B}. On the left, we set the phases $\chi = 0$ and show: in dashed, the cases with fixed $\zeta_1+\zeta_2 = 0.2$; in dashed-dotted, a case with a different $\zeta_1+\zeta_2$. In the right panel, we set $\chi \neq 0$ and $\zeta_1 = \zeta_2 = 0.1$. The expanded expressions of Eq.~\ref{eq:Muellerx} with only the $\zeta$ and $\chi$ different from zero are the following:
	\begin{equation} \label{eq:Muellerx_z}
		\begin{split}
			M^{TT}_{x} & \simeq \frac12 + \frac12 \left(\zeta_1 \cos\chi_1  -\zeta_2 \cos\chi_2 \right)\sin(2\theta),  \\ 
			M^{TQ}_{x} & \simeq \frac12 \cos(4 \theta) -\frac12 \left(\zeta_1  \cos\chi_1-\zeta_2  \cos\chi_2 \right)\sin (2\theta)-\frac12\left(\zeta_1  \cos\chi_1 +\zeta_2  \cos\chi_2 \right) \sin (4\theta),\\
			M^{TU}_{x} & \simeq \frac12 \sin(4 \theta) + \frac12 \left(\zeta_1 \cos\chi_1 -\zeta_2 \cos\chi_2 \right)\cos (2\theta)+\frac12 \left(\zeta_1 \cos\chi_1 +\zeta_2 \cos\chi_2\right) \cos (4\theta) .
		\end{split}
	\end{equation}
	
	In both bottom panels of Fig.~\ref{fig:plot_id_B}, we can appreciate the effect of cross-polarization: the shape of the $BB$ spectrum is modified by the $E \rightarrow B$ leakage, enhancing the final spectrum. This should be compared with the effect of $h$ and $\beta$, which instead act to rescale the input $BB$ spectrum. For this reason, a value of $\zeta$ of the same order of magnitude of $h$ causes a more prominent effect on the final spectra. Of course, the effect is stronger in combination with $\zeta_{1,2}$.
	In the left panel, we can see that the two dashed lines, corresponding to two different choices of $\zeta_1-\zeta_2$, overlap: the $2 \theta$ terms, canceled out by the map-making procedure,
	don't impact on the $BB$ spectrum. In the right panel, the three curves are similar but not perfectly overlapping, as we are varying the sum $\left(\zeta_1 \cos\chi_1 +\zeta_2 \cos\chi_2\right)$ by changing the phases of $\chi$. We note that the difference between the curves is mainly driven by the high value of $\zeta_{1,2}$. At first order, the  $\chi$ phases act as a small real multiplicative factor (see Eq. \ref{eq:Muellerx_z}),
	so perturbing them has the same effect of slightly changing the module of $\zeta$. For that reason, in the following, we act only on $\zeta$ and keep $\chi = \chi_s = 0$.
	
	\subsubsection{Non-ideal mapmaking matrix $B$}
\begin{figure*} 
		\centering
		{\includegraphics[width=17 cm]{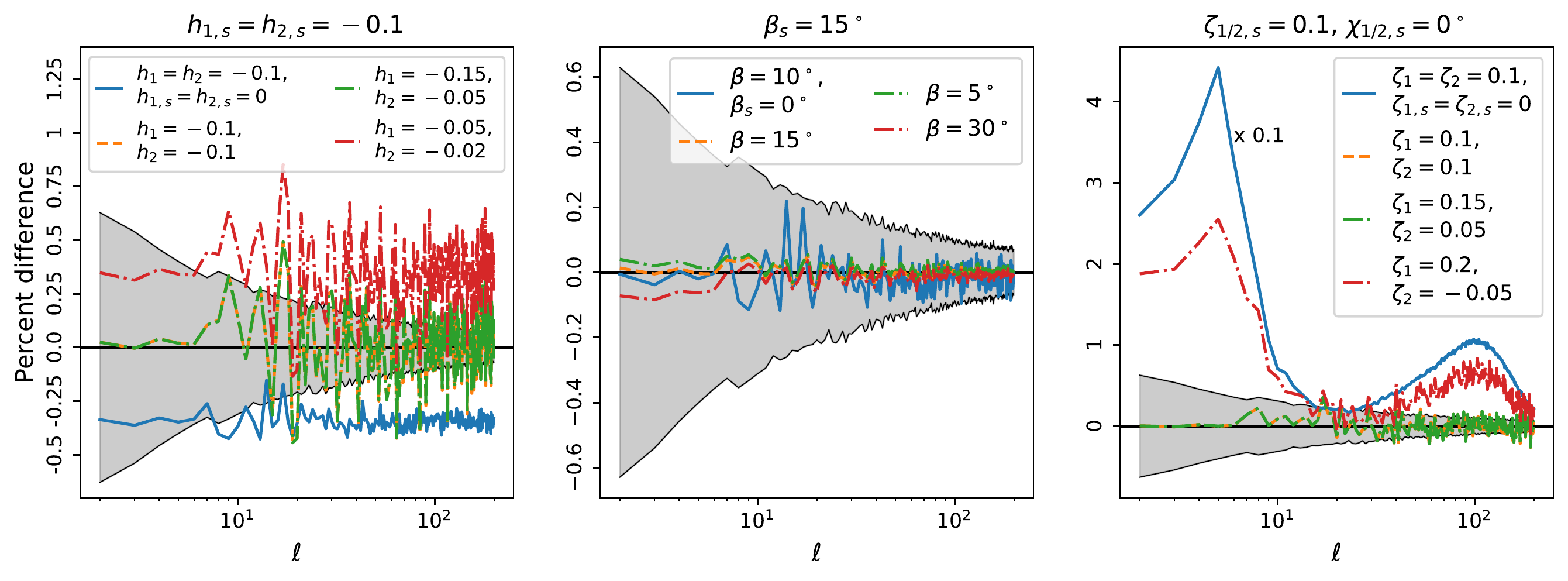}}
		\caption{Percent difference of the noise de-biased $C^{BB}_{\ell}$(CMB + noise + systematics) 
			with respect to the ideal $C_{\ell}^{BB}$(CMB only), when assuming a non-ideal mapmaking matrix $B$. The shaded gray region shows the standard deviation of the ideal CMB realizations, normalized to their mean. The parameters with and without the subscript $s$ enter the mapmaking matrix $B$/TOD matrix $A$. In each panel, we fix the value of each class of solver parameters (indicated in the corresponding title), except for the solid blue curve, showing (for reference) the case with non-ideal pointing matrix $A$ and ideal mapmaking matrix $B$.  In orange dashed lines, we report the results in the case $A = B$. In the dashed-dotted lines, we report the results in the case $A \neq B$ (different colors correspond to different values of the HWP parameters, see legend). In the right-most panel, the case with ideal $B$ (blue solid) is divided by a factor of 10 with respect to the actual signal to ease the comparison with the other curves. The lines are wiggly because of the noise de-biasing.  See the text for discussion.
		}\label{fig:plot_non_id_B}
	\end{figure*}

	In Fig.~\ref{fig:plot_non_id_B}, we report the results for the study with a non-ideal mapmaking matrix. We plot the percent difference of the average over 100 $BB$ (noise de-biased) spectra affected by systematics with respect to the average over the spectra from the same CMB realizations, not affected by systematics.
	
	In this case, we use the following values for the parameters in the solver matrix $B$ (indicated with the subscript $s$): $h_{1,s} = h_{2,s} = -0.1,\,\beta_s=\zeta_s=0$ (left panel), $\beta_s = 15^\circ,\,h_s=\zeta_s=0$ (middle panel), $\zeta_{1,s} = \zeta_{2,s} = 0.1,\,\beta_s=h_s=0$ (right panel). Results corresponding to this choice are shown in Fig.~\ref{fig:plot_non_id_B}, using dashed and dashed-dotted lines. For reference, we also include the residual spectra obtained when assuming the ideal $B$ (blue solid). 
	We want to stress that since the map-making matrix $B$ is different between the panels, also the noise bias is slightly different (see Sect.~\ref{sec:mono_setup}). 
	
	
	Different cases are shown in Fig.~\ref{fig:plot_non_id_B}:
	\begin{itemize}
		\item[$\bullet$]  $A = B$, that is, TOD HWP equal to map-making HWP 
		(orange dashed line in all panels), which would perfectly correct for systematic effects in a noiseless case. The correction is less visible in our case due to noise;
		\item[$\bullet$] for $x\equiv h,\zeta$, the green (red) dashed-dotted line in the leftmost panel corresponds to $x_1+x_2=x_{1,s}+x_{2,s}$ ($x_1+x_2\neq x_{1,s}+x_{2,s}$). We note that the green line overlaps with the orange line, as expected in the case $A=B$;
		\item[$\bullet$] for $\beta$, the green (red) dashed-dotted line in the middle panel has $\beta < \beta_s$ ($\beta >\beta_s$), which gives a slightly positive (negative) shift. In fact, we would expect a correction of order $\cos \beta_s$, while a higher (smaller) $\cos \beta$ enters in the TOD matrix;
		\item[$\bullet$] in the case of $\zeta$ (rightmost panel), having a mismatch of the type: $\zeta_1+\zeta_2 \neq \zeta_{1,s}+\zeta_{2,s} $ always causes a positive bias because it provides $E \rightarrow B$ leakage.
	\end{itemize} 
	It is interesting to observe that the red dashed-dotted line for $h$, which refers to $|h_1+h_2| < |h_{1,s}+h_{2,s}|$, corresponds to an overall shift of the spectrum toward higher values, as we are over-correcting for $h$.
	The differences between the ideal and realistic $C_{\ell}^{BB}$ can be quantified by estimating the corresponding bias on the tensor-to-scalar ratio $r$ (see Sect.~\ref{subsec:deltar} for detail). However, we defer this detailed discussion to the more realistic multi-frequency analysis, in the following Sections.
	Nevertheless, as a qualitative prediction in this simple setting, we would expect, in general, a negative $\Delta r$ for $|h| > |h_s|$, a positive (negative) $\Delta r$ for $\cos \beta > (<) \cos \beta_s$, and a positive $\Delta r$ for $\zeta \neq \zeta_s$.

	\section{Multi-frequency case}\label{sec:cromo}
	In this section, we consider a more realistic scenario. In particular, we move from the Dirac-delta response of the detectors in frequency as implicitly assumed in the previous section to a top-hat response. This allows us to take into account two main effects that were previously neglected. First of all, we allowed for HWP matrix elements to be frequency-dependent: we assumed a specific LiteBIRD MHWP design with certain frequency profiles within each band. The MHWP performance was computed using realistic models developed for previous waveplate applications~\citep{pisano2020development}. 
	Secondly, we included a frequency-dependent foreground component in the input sky maps. We describe below the details regarding the inclusion of the two new effects. The analysis follows the same steps detailed in Sect.~\ref{sec:strategy}. 
	
	\subsection{Setup for the multi-frequency analysis} \label{subsec:setup}
	We considered four frequency bands corresponding to the MFT channels of the proposed LiteBIRD satellite~\citep{Hazumi:2021yqq} and one frequency band each for LFT and HFT, the closest ones to the CMB channels. The central frequency, band-width, and FWHM of the Gaussian beam for each channel are summarized in Table~\ref{tab:bands}. 
	
	\begin{table}
		\begin{center}  
			\caption{LiteBIRD bands used for the multifrequency analysis.}  \label{tab:bands}    
			\begin{tabular}{p{1.3cm}|p{1.4cm}|p{1.4cm}|p{1.3cm}|p{1.3cm}}
				\hline
				\small{Telescope} & \small{Central frequency [GHz]} & \small{Bandwidth (frac.) [GHz]} & \small{Gaussian beam size [arcmin]} & \small{NET$_{array}$ [$\mathrm{\mu K_{CMB} \sqrt{s}}$]} \\
				\hline
				LFT & $100$ & 23 (0.23)  & 30.2 & 5.11\\
				\hline
				MFT & $100$ & 23 (0.23)  & 37.8 &  4.19\\
				MFT & $119$ & 36 (0.30)  &  33.6 & 2.82\\
				MFT & $140$ & 42 (0.30)  & 30.8 & 3.16 \\
				MFT & $166$ & 50 (0.30)  & 28.9 & 2.75\\
				\hline
				HFT & $195$ & 59 (0.30)  & 28.6 & 5.19\\
				\hline
			\end{tabular}
		\end{center}
	\end{table}
	
	The input sky is different from that used in Sect.~\ref{sec:mono}. We only considered one CMB realization from the fiducial spectra chosen by LiteBIRD collaboration, (in prep.), with $r = 0$ and $\tau =0.0544$. We go on to explain the reasoning behind this later on in this work. We decided to work in $\mathrm{\mu K_{CMB}}$ units, so that the CMB signal is independent from the frequency in the frequency range that we consider in this work
	\footnote{In the radio-domain, it is customary to express the brightness (emitted intensity) $I(\nu)$ at a given frequency $\nu$ as the brightness of a black-body $BB_\nu(T_b)$ with temperature $T_b$ at the same frequency: $I(\nu)=BB_\nu(T_b)$. $T_b$ is the \textit{brightness temperature}. In the Rayleigh-Jeans (RJ) regime ($h\nu\ll k_BT$), we can take a Taylor expansion around $h\nu/kT$ so that $I(\nu)\simeq (2k_B\nu^2/c^2) T_\mathrm{RJ}$.\\CMB maps are usually given in units of linearized differential temperature (maps show fluctuations around the CMB mean temperature $T_\mathrm{CMB}$): $dI(\nu)=(dBB_\nu(T_b)/dT_b) dT_b\rightarrow \Delta T_\mathrm{CMB}=\Delta I(\nu)/(dBB_\nu(T)/dT)|_{T=T_\mathrm{CMB}}$. Using the same linearized expression, the brightness in RJ units is given by: $\Delta T_\mathrm{RJ}=\Delta I(\nu)/(dBB_{\nu(T),RJ}/dT)$, where $BB_{\nu(T),RJ}$ is the Taylor-expanded black-body emission in the RJ regime. Commonly RJ units are used for foreground emissions \citep{planck2014-a12,planck2016-l04}.\label{fn:units}}.
	
	We added frequency-dependent maps of foreground emissions to the CMB maps. To do so, we generated a set of foreground maps for the range of frequency used in this work. Foreground maps are generated with the \texttt{PySM} software package~\citep{Thorne_2017}. We adopted the [d1,s1,a1,f1] model available in \texttt{PySM}, with spatially varying spectral indices of dust and synchrotron. Both the CMB and foreground maps have a resolution of $N_\mathrm{side}=512$ and are smoothed with a Gaussian beam with FWHM given by the LiteBIRD resolution at the given frequency channel (see Table~\ref{tab:bands}). To keep things simple and to set the focus on the possible chromaticity of systematic effects, in this multi-frequency analysis, we neglected the contribution of instrumental noise. The signal, $d_\mathrm{obs}(t_i,p,\nu),$ observed at a given time sample, $t_i$, in a certain pixel, $p,$ in a given frequency channel, $\nu,$ is simply the weighted average of the sky signal in that frequency band. Weights are given by the frequency-dependent Mueller matrix elements, $M_\mathrm{full}(\nu),$ of the optical system.
	
\begin{figure}[!htbp] 
		\centering
		\includegraphics[width=0.5\textwidth]{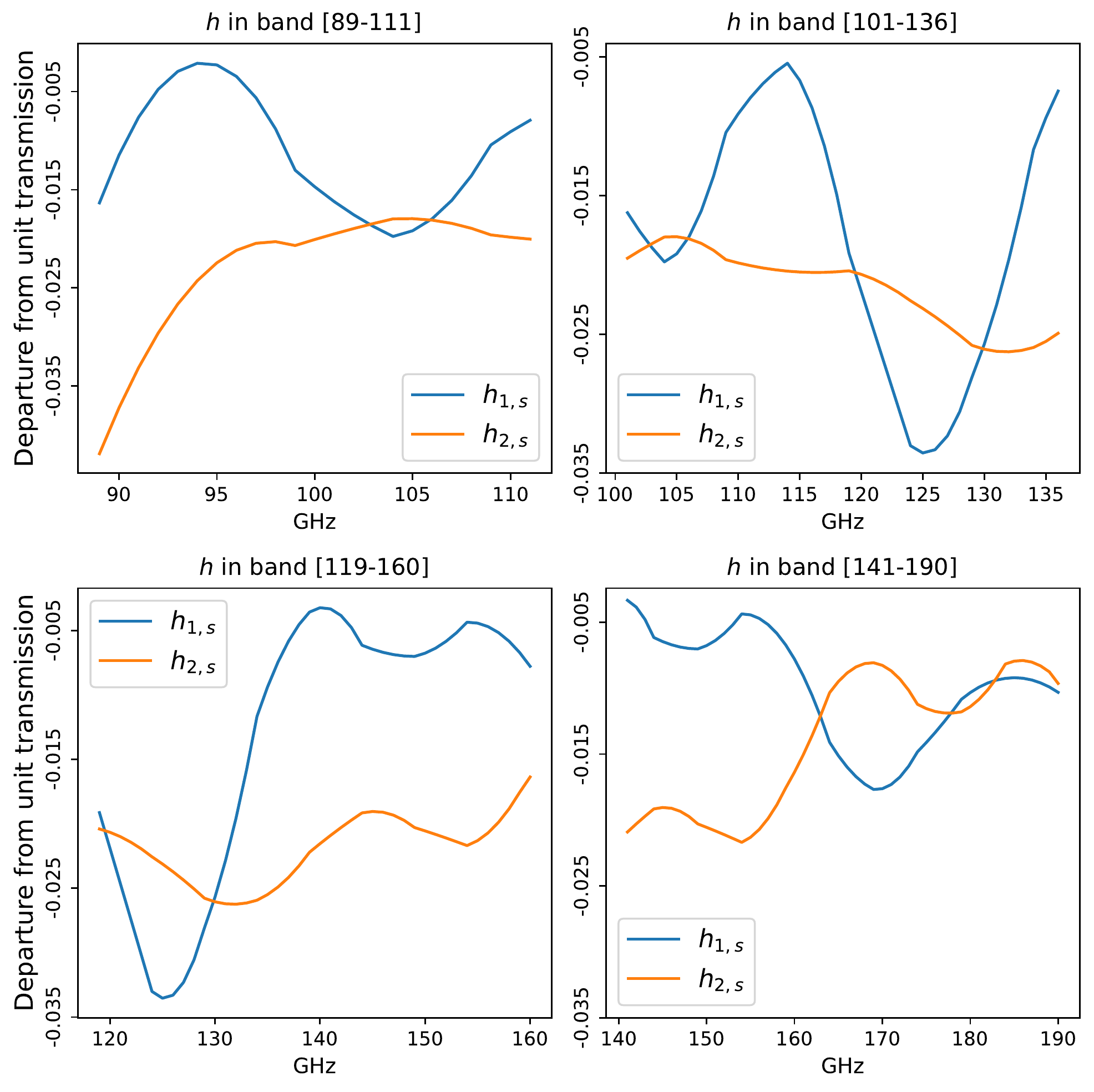}
		\caption{Simulated profiles of the MFT HWP transmissions $h_1,h_2$, for the four selected MFT frequency bands. The subscript $s$ indicates that they are used in the solver matrix $B$.}\label{fig:prof_h}
	\end{figure}

\begin{figure}[!htbp]
		\centering
		\includegraphics[width=0.5\textwidth]{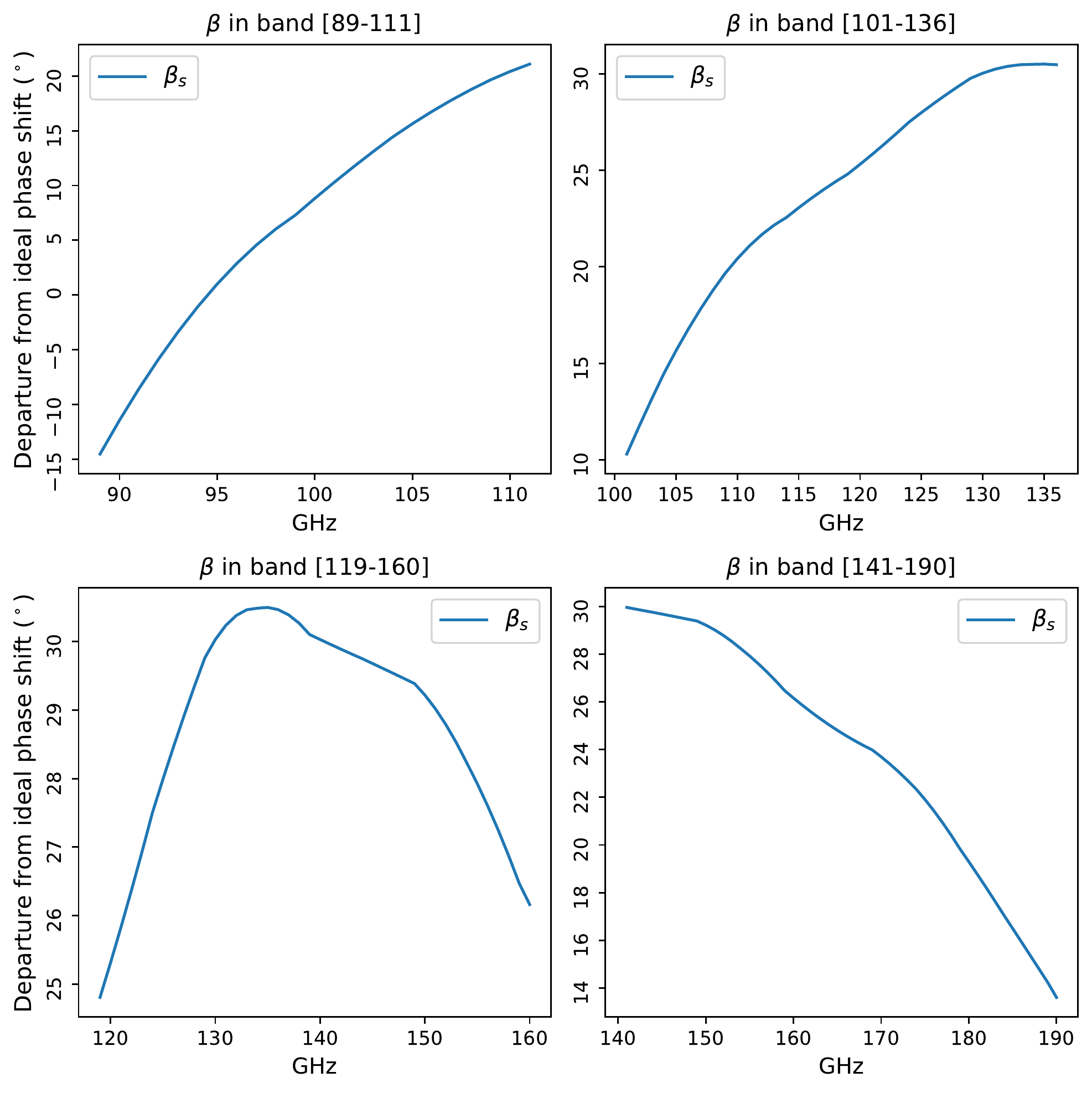}
		\caption{Simulated profiles of the HWP phase-shift $\beta$, for the four selected MFT frequency bands of LiteBIRD. The subscript $s$ indicates that they are used in the solver matrix $B$.} \label{fig:prof_b}
	\end{figure}

	In general, HWP parameters show a non-trivial dependence on the frequency of the incident signal due to fabrication details. Finite-element modeling and a laboratory characterization of the HWP devices allows reconstruction of the expected profiles of the HWP matrix elements~\citep{Pisano:2014eba,pisano2020development,Pisano:crosspolcite}. 
	In this analysis, we adopted a model of the HWP derived for the MFT channels \citep[see Figures \ref{fig:prof_h} and \ref{fig:prof_b};][]{Montier:2021tjz,lamagna2021optical}. 
	In a simulated Mesh-HWP the cross polarization parameters $\zeta$ are exactly null because of the supposed exact symmetry of the system. Because of this, we are not able to access the frequency profiles of $\zeta_1, \zeta_2$ and so, for simplicity, we assumed them to be constant and equal to $10^{-2}$. In reality, the symmetry of the configuration could be spoiled and $\zeta$ could be as large as the value we consider. Also for the L/HFT bands we take constant values for all the systematic parameters (for LFT: $h = -0.015, \beta = 7.19^\circ, \zeta = 0.01$, for HFT: $h = -0.01, \beta = 15^\circ,\zeta = 0.01$). We checked that this does not affect significantly the final result (see Appendix \ref{app:flat_prof}). The solver matrix $B(\nu)$ is built on this model. The pointing matrix $A(\nu)$ is a perturbed version of the same model, as we explain later in the text. The Mueller matrix elements of the realistic HWP as given by our model are shown in Fig.~\ref{fig:prof_M} for each frequency channel. 
	
	We are thus ready to express the TOD sample $d_\mathrm{obs}(t_i,p,\nu)$ as a generalized version of Eq. \ref{eq:dobsti}, splitting the contribution from CMB (in CMB units) and from the foregrounds (FG, in Rayleigh-Jeans units):
	
	
	\begin{equation}\label{eq:dobsnu}
		\begin{split} d_\mathrm{obs}(t_i,p,\nu)=
			&\frac{\int d\nu F_{CMB}(\nu) \left[
				M^{TT}_{i}(\nu) T_{CMB}(p)+
				M^{TQ}_{i}(\nu) Q_{CMB}(p)+
				M^{TU}_{i}(\nu) U_{CMB}(p) \right]}{\int d\nu F_{CMB}(\nu) }\\
			&+\frac{\int d\nu F_{FG}(\nu) \left[
				M^{TT}_{i}(\nu) T_{FG}(\nu,p)+
				M^{TQ}_{i}(\nu) Q_{FG}(\nu,p)+
				M^{TU}_{i}(\nu) U_{FG}(\nu,p) \right]}{\int d\nu F_{CMB}(\nu)}.
		\end{split}
	\end{equation}
	The terms $F_{CMB,FG}$ in Eq. \ref{eq:dobsnu} are given by: 
	\begin{subequations}
		\begin{align}
			F_{CMB}(\nu) &= \frac{\partial BB(\nu,T)}{\partial T_{CMB}}\tau_c(\nu), \\
			F_{FG}(\nu) &= \frac{\partial BB_{RJ}(\nu,T)}{\partial T_{RJ}}\tau_c(\nu).
		\end{align}
	\end{subequations}
	where $\tau_c(\nu)$ is the top-hat bandpass, including the throughput factor, $\nu^2$, and the black-body derivatives take into account the conversion from CMB and RJ units, respectively (see footnote~\ref{fn:units}). The common denominator in Eq. \ref{eq:dobsnu} sets the final units as $\mathrm{\mu K_{CMB}}$. We note that in Eq. \ref{eq:dobsnu} $T_\mathrm{CMB}, Q_\mathrm{CMB}, U_\mathrm{CMB}$ are independent of frequency and may, therefore, be taken out from the first integral. In matrix form, we can express the following 
	
	\begin{equation}\label{eq:tod_nu}
		\mathrm{d}_\mathrm{obs}(\nu)=A_\mathrm{CMB} \, m_\mathrm{CMB} + \int  A_\mathrm{FG}(\nu) \, m_\mathrm{FG}(\nu) d\nu,
	\end{equation}
	
	\noindent where $m_\mathrm{CMB}$ is the (3N$_{\rm pixels}$) vector of CMB Stokes parameters and $m_\mathrm{FG}(\nu)$ is the (3N$_{\rm pixels}$) vector of foreground Stokes parameters as a function of frequency. The pointing matrix $A_\mathrm{CMB}$ is a frequency-independent (3N$_{\rm pixels} \times$ N$_{\rm samples}$) matrix with elements given by the first integral in Eq.~\ref{eq:dobsnu} over the Mueller elements\footnote{Both $A_{CMB,i}$ and $A_{FG,i}$ are non-zero elements of the  relative pointing matrices corresponding to the pixel $p_i$.}
	
	\begin{equation}\label{eq:Acmbnu}
		A_{CMB,i} = \left( \frac{\int d\nu F_{CMB}(\nu) {\color{red}{M^{TT}_{i}(\nu)}} }{\int d\nu F_{CMB}(\nu)}, \frac{\int d\nu F_{CMB}(\nu) {\color{red}{M^{TQ}_{i}(\nu)}} }{\int d\nu F_{CMB}(\nu)}, \frac{\int d\nu F_{CMB}(\nu){\color{red}{M^{TU}_{i}(\nu)}} }{\int d\nu F_{CMB}(\nu)} \right).
	\end{equation}
	The subscript $i$ indicates the time sample, as before. The pointing matrix $A_\mathrm{FG}(\nu)$ is a frequency-dependent (3N$_{\rm pixels} \times$ N$_{\rm samples}$) matrix with elements given by the Mueller elements in the second integrand in Eq.~\ref{eq:dobsnu}:
	
	\begin{equation}\label{eq:Afgnu}
		A_{FG,i}(\nu) = \left( \frac{F_{FG}(\nu) {\color{red}{M^{TT}_{i}(\nu)}} }{\int d\nu F_{CMB}(\nu)}, \frac{F_{FG}(\nu) {\color{red}{M^{TQ}_{i}(\nu)}} }{\int d\nu F_{CMB}(\nu)}, \frac{F_{FG}(\nu) {\color{red}{M^{TU}_{i}(\nu)}} }{\int d\nu F_{CMB}(\nu)} \right).
	\end{equation}
	
\begin{figure}[!htbp]
		\centering 
		\includegraphics[width=0.5\textwidth]{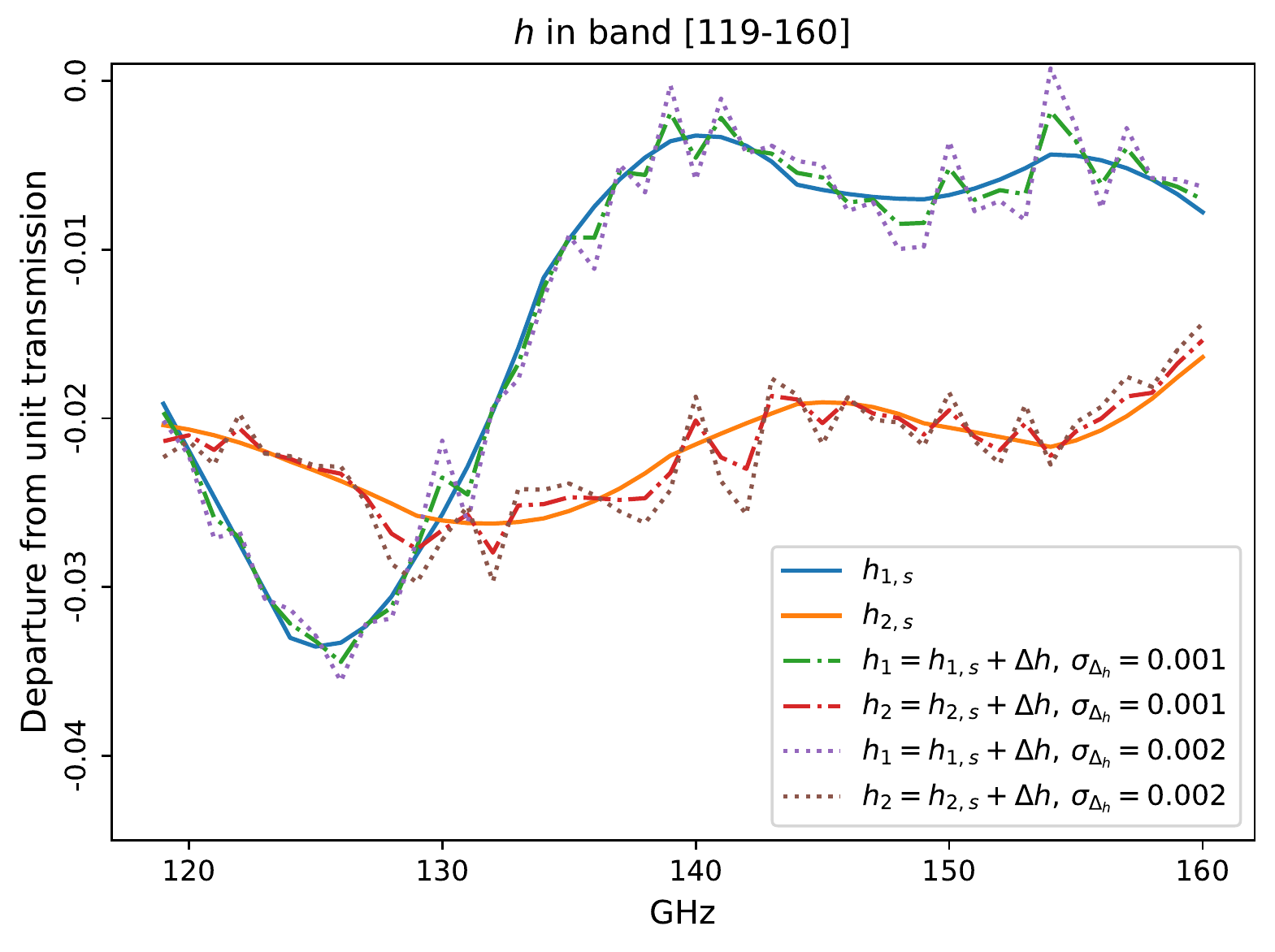}\quad
		\caption{Example of perturbations of the HWP profiles: Dashed-dotted lines show one realization of perturbed profiles for $h_1, h_2$ with $\sigma_{\Delta h} = 0.001$, while the dotted lines show the same realization with higher $\sigma_{\Delta h} = 0.002$.} \label{fig:prof_pert}
	\end{figure}

	The mapmaking procedure then consists of inverting Eq. \ref{eq:tod_nu} along the same lines as described in Sect.~\ref{sec:mono}. The estimated map is given by: 
	
	
	\begin{equation} \label{eq:m_out}
		\begin{split}
			&m_{out} = N^{-1} \left( \sum_i B_i^T  d_\mathrm{obs}(t_i) \right) =  N^{-1} \left( \sum_i B_i^T A_\mathrm{CMB,i} \, m_\mathrm{CMB}\right) + N^{-1} \left(\sum_i B_i^T \int  A_\mathrm{FG,i}(\nu) \, m_\mathrm{FG}(\nu) d\nu \right),
		\end{split}
	\end{equation}
	where
	\[
	N^{-1} = \left(\sum_i B_i^T B_i \right)^{-1}
	\]
	and again $B$ is the solver matrix integrated over frequency. To build the solver matrix in the multi-frequency case, we adopt the following procedure. We define $B$ as
	
	\begin{equation}\label{eq:Bnu}
		\begin{split}
			B_i &= \left( \frac{\int d\nu F_{CMB}(\nu) {\color{red}{M^{TT}_{s,i}(\nu)}} }{\int d\nu F_{CMB}(\nu)}, \frac{\int d\nu F_{CMB}(\nu){\color{red}{M^{TQ}_{s,i}(\nu)}} }{\int d\nu F_{CMB}(\nu)},\frac{\int d\nu F_{CMB}(\nu){\color{red}{M^{TU}_{s,i}(\nu)}} }{\int d\nu F_{CMB}(\nu)}\right), 
		\end{split}
	\end{equation}
	where $M^{TT}_{s,i}(\nu),\,M^{TQ}_{s,i}(\nu),\,M^{TU}_{s,i}(\nu)$ contains our model profiles of the frequency-dependent Mueller matrix elements in Eqs.~\ref{eq:Acmbnu}-\ref{eq:Afgnu}. As before, the subscript $i$ refers to the time samples. The subscript $s$ labels the matrices entering the map-making matrix $B$, in parallel with the solver parameters defined in Sect. \ref{sec:mono_setup}. The profiles for $h$ and $\beta$ are shown in Figs.~\ref{fig:prof_h} and~\ref{fig:prof_b}. In practice, the frequency profile  $x(\nu)$ of the HWP parameter $x=h,\zeta,\beta$ is used to build the matrix $B$ in Eq.~\ref{eq:Bnu}, while the perturbed profile $x(\nu)+\Delta x(\nu)$ is used to build the matrices  $A_\mathrm{CMB}$ and $A_\mathrm{FG}$ in Eqs.~\ref{eq:Acmbnu}-\ref{eq:Afgnu}. The perturbation $\Delta x(\nu)$ is treated as a Gaussian fluctuation around the ``true'' value of $x(\nu)$ with a variance of $\sigma_{\Delta x}$, and can therefore be either positive or negative. An example of the perturbed profiles is shown in Fig.~\ref{fig:prof_pert}. 
	
	We note that for each parameter, we considered uncorrelated perturbations in the frequency ($\langle\Delta x(\nu) \Delta x(\nu')\rangle = \delta_{\nu\nu'} \sigma^2_{\Delta x}$). In addition, perturbations in one parameter are also uncorrelated with perturbations in a different parameter ($\langle\Delta x(\nu) \Delta y(\nu)\rangle = \delta_{xy} \sigma^2_{\Delta x}$). With this treatment, we aim to simulate the realistic, albeit simplified case of having a mismatch between the model for the profile of each systematic to be used in the solver matrix $B$ as well as in the profiles actually entering in the TOD matrix $A$. This mismatch is, for example, produced by calibration errors on the parameters. Modeling more realistic error distributions would require a clear knowledge of the optical chain and calibration setup used to perform laboratory measurements. Since this is not available at the current stage, we prefer to defer this topic to a detailed study to future publications. Notwithstanding, the simplified modeling employed in this work remains a valid approach for the pedagogical purposes of this analysis. It is worth noting that even in the ideal case of $B=A_{CMB}$, that is, in a perfect calibration of the HWP profiles, the HWP non-idealities coupled with the mapmaking procedure intrinsically produce a distortion of the foreground field that needs to be deprojected~\citep[see Sect.~\ref{subsec:depr};][]{Verges_Errard_Stompor}.
	
	For each non-ideal parameter $h,\zeta,\beta$, we considered different choices of $\sigma_{\Delta x}$, as summarized in Table~\ref{tab:sigmax}. To avoid  oversensitivity to the specific realization of the pointing matrix $A$, we ran ten simulations for each systematic parameter and each band, resulting in ten realizations of the pointing matrix $A$. This way, each value of $\sigma_{\Delta x}$ can  share the same realizations and the comparison between cases with different $\sigma_{\Delta x}$ thus depends just on the amplitude of the error (within the same band and systematic parameter). To keep matters simple, we used one CMB and FG realization, that is, we kept a fixed input sky. In fact, the residuals coming from systematic effects are mostly dominated by the much brighter foregrounds and not particularly affected by the actual CMB realization. The choice of keeping the same foreground model in the analysis is considered further in Sect.~\ref{subsec:depr}.

	\begin{table}
		\begin{center}  
			\caption{List of the errors in each HWP systematic parameter per unit frequency resolution}   \label{tab:sigmax}        
			\begin{tabular}{ p{2.3 cm}|p{1.5cm}|p{1.5cm}|p{1.5cm}|p{1.5cm}|p{0.7cm}}
				\hline
				$\sigma_{\Delta h} \, \left[\sqrt{\text{GHz}}\right]$ & 0.001 & 0.002 & 0.003 & 0.005 & \\ 
				\hline
				$\sigma_{\Delta \beta} \, \left[^\circ \, \sqrt{\text{GHz}}\right]$ & 0.5 & 1 & 2 & 3 & 5 \\ 
				\hline
				$\sigma_{\Delta \zeta} \, \left[\sqrt{\text{GHz}}\right]$ & 0.001 & 0.002 & 0.0035 & 0.005 & \\ 
				\hline
			\end{tabular} 
		\end{center}
		{\raggedright \textbf{Notes.} Assumed to be random in frequency, those errors scale with the frequency resolution $\Delta \nu$ like $1/\sqrt{\Delta \nu}$. The values of $\sigma$ for each kind of parameter are chosen as a fraction of the average values of $h$ and $\beta$, from the simulated profiles, and of the selected level of $\zeta = 10^{-2}$. \par}
	\end{table}
	
	As done in Sect.~\ref{sec:mono}, we vary one parameter at a time. In addition, we also consider the case of joint variation of multiple parameters to test for possible correlations between systematic effects induced by HWP non-idealities. 
	
	With regard to the units of $\Delta\sigma_x$ in Table \ref{tab:sigmax}, we note that  in this work, we assume that we are able to reconstruct the HWP profiles with a resolution of $\Delta\nu=1\,\mathrm{GHz}$, so that $\Delta\sigma_x$ effectively refers to the accuracy on the $x$-parameter per unit resolution. Had a different frequency resolution been chosen, the corresponding accuracy would be scaled as $1/\sqrt{\Delta\nu}$. 
	
	\subsection{Deprojection template}\label{subsec:depr}
	In Sect.~\ref{sec:mono}, residuals were shown with respect to the ideal case of perfect knowledge of the optical system. The reason was that we wanted to focus on the physical effects induced by HWP non-idealities to the observed quantities. In this section, we follow a closer approach to that can be applied to realistic observations. Residuals are shown with respect to a template of our best estimate of the input sky. To build this template, we assume an input CMB sky $m_{CMB}$ and an input FG model $m_{FG}(\nu)$. We then simulate observations of CMB+FG following the same algorithm described above for multi-frequency observations, with the only difference that we employ the matrix $B$ both as the pointing matrix and the solver matrix. At the end of the mapmaking procedure, we are left with the template:
	
	\begin{equation}\label{eq:template}
		\begin{split}
			&m_{\rm templ} = N^{-1} \left( \sum_i B_i^T B_{i} \, m_{CMB}\right) + N^{-1} \left(\sum_i B_i^T \int  B_{FG,i}(\nu) \, m_{FG}(\nu) d\nu \right)= m_{CMB} + N^{-1} \left(\sum_i B_i^T \int  B_{FG,i}(\nu) \, m_{FG}(\nu) d\nu \right) ,
		\end{split}
	\end{equation}
	
	\noindent
	where $B_{FG,i}(\nu)$ is the matrix in Eq.~\ref{eq:Afgnu} with $M^{TX}_i(\nu)=M^{TX}_{s,i}(\nu)$.
	
	The residuals $\mathcal{R}$ are then computed as the difference between the output maps (Eq.~\ref{eq:m_out}) and the template, $\mathcal{R}\equiv m_\mathrm{out}-m_\mathrm{templ}$. 
	The template $m_\mathrm{templ}$ is built with the same CMB map and the same FG model used as an input to generate the output maps $m_\mathrm{out}$. In principle, we could have made a different choice, as, indeed, in the real case, it may well be that the estimated sky used to generate a template does not perfectly match the ``true'' observed sky. This would of course be the cause of differences between the output maps and the template map, even if instrumental systematics were perfectly corrected for. However, it has been proved that an iterative approach can be employed~\citep{planck2016-l03,Delouis:2019bub} that converges quickly to the best template estimate, even when the initial sky models are much different from the observed sky. Therefore, in this work, we use the same sky model to generate both the deprojection template and the output maps. In doing so, we rather focus on possible differences between the output maps and the template that are only due to unaccounted HWP systematics and neglect the coupling of component separation with systematics. Here, we include the uncertainties from component separation individually in the noise term that we add to the power spectrum (see Sect. \ref{subsec:deltar}). By construction, the residual due to  unaccounted-for systematics is null if $A_{CMB}=B$ (TOD HWP = map-making HWP). This requires measuring the systematic parameters with sufficient precision. 
	
	We want to stress that the deprojection procedure does not only reduce the residual systematics in case of imperfect knowledge of the HWP profiles, but it also corrects for the intrinsic foreground distortion due to the frequency-dependent non-idealities. Indeed, as is evident in making the comparison between Eqs. \ref{eq:Acmbnu}, \ref{eq:Afgnu}, and \ref{eq:Bnu}, the mapmaking procedure allows for the recovery of unbiased estimates of the CMB component. However, the same is not true for the foregrounds, even if $A_{CMB}=B$. The frequency dependence of the HWP parameters introduces a 
	band integration for each sample that not only depends on the direction of observation (as in a usual band-pass integration), but also on the rotation angle of the HWP. 
	This peculiarity of band integration in the presence of a frequency-dependent HWP is particularly dangerous, requiring not only an accurate optical characterization but also HWPs with as flat as possible in-band properties~\citep{Verges_Errard_Stompor}.
	
	Finally, we would like to note that the template-subtraction procedure is not
	intended to serve as a component separation technique. Rather, this procedure has to be intended as a deprojection algorithm that allows for the isolation of the propagation of systematics in the output (or, in real scenarios, observed) maps. As such, this technique has been already proven to be efficient and has been employed in data analysis pipelines for recent CMB experiments~\citep{planck2016-l03, Ade:2015fpw,Delouis:2019bub}. From the definition of the residuals, $\mathcal{R}$, we can easily see that in our case, a non-vanishing residual is clearly due to the mismatch between the pointing and the solver matrices.
	
	\subsection{Results of the multi-frequency analysis} \label{sec:results}
	The residual power spectra caused by perturbations in each systematic parameter are presented here. The residual maps, $\mathcal{R}=m_\mathrm{out}-m_\mathrm{templ}$, computed in each frequency band are masked with a $f_{sky}=70\%$ galactic mask $M_{70}$, previously apodized with 5$^\circ$ apodization scale. Then the power spectra are simply computed with the \texttt{healpy} routine \texttt{anafast}, correcting for the sky fraction, $f_{sky}$, and the beam window function, $b_{\ell}$:
	\begin{equation}\label{eq:resSpectra}
		C_{\ell}^{res} = C_{\ell}(\mathcal{R} \times M_{70})/(f_{sky} \, b^2_{\ell}).
	\end{equation}
	
	We checked that the use of more sophisticated power spectrum estimators \citep{Alonso_2019} does not make a significant difference in our analysis. 
	Indeed, the residuals are dominated by the foregrounds. The foreground emission in $EE$ and $BB$ is of the same order of magnitude. The $E$-to-$B$ leakage due to partial sky coverage (not corrected for when using the \texttt{anafast} estimator) is therefore much weaker than what would be expected in the case of a CMB-only signal. As a result, we can safely neglect the $EE-BB$ mixing when using the \texttt{anafast} estimator.
	Furthermore, when applying $E-B$ purification techniques, the estimator may still be biased if the mask apodization procedure is not optimized individually in each multipole bin~\citep{Ferte:2013, Ferte:2015}. Performing this optimization is beyond the scope of this paper, so we decided to employ the simplest estimator available.
	
\begin{figure*} 
		\centering 
		\includegraphics[width= 17 cm]{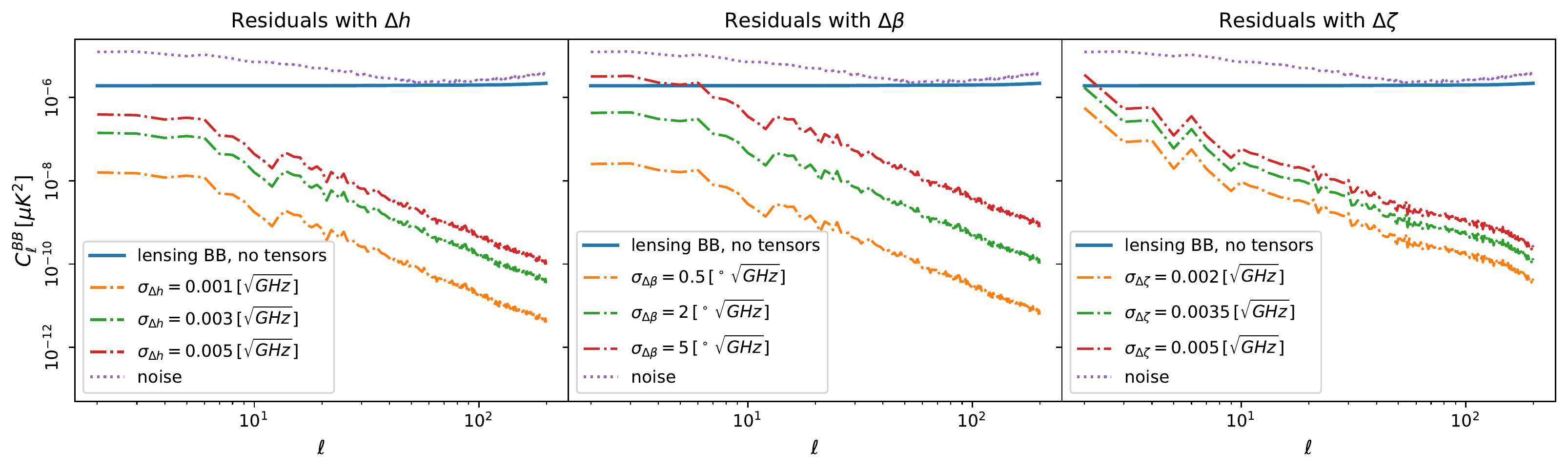}\quad
		\caption{Residual $BB$ power spectra for the frequency band centered at 140 GHz. The thick blue lines are the fiducial $C_{\ell}^{BB}$(CMB) (lensing $BB$, $r = 0$). The dashed-dotted lines show the residual spectra $C_{\ell}^{BB,res}$ obtained as the average over ten Gaussian error realizations of the perturbations applied to each systematic parameter. Different colors correspond to different values of the variance $\sigma_{\Delta x}$ of the error realizations applied to the parameter $x$. The dotted line is the noise bias $C_{\ell}^{BB,noise}$, as described in Sect.~\ref{subsec:deltar}.} \label{fig:res}
	\end{figure*}

	In Fig.~\ref{fig:res}, we report an example of the $BB$ residual power spectra for the $140\,\mathrm{GHz}$ frequency band. The residuals clearly follow a power-law behavior. This is expected, since most of the residual power due to the mismatch between the pointing matrix and the mapmaking matrix comes from foregrounds. As already explained in Sect.~\ref{subsec:setup}, we computed ten realizations of the perturbed profiles for each systematic parameter in each frequency band with the values of $\sigma_{\Delta x}$ in Table~\ref{tab:sigmax}. We finally computed the residual power spectrum for each $\sigma_{\Delta x}$ as the average of the ten spectra corresponding to each error realization.
	
	We also compute the coaddition of residual maps from different channels, resembling a rough component separation procedure. We implemented it as the weighted average of the maps from each channel:
	\begin{equation}\label{eq:weight}
		m^x_{res,tot} = \frac{\sum_i m^x_{res,i} w_i}{\sum_i w_i},
	\end{equation}
	where $x = \{h,\beta,\zeta\}$, the sum $i$ runs over the frequency channels, 
	and $w_i$ is the corresponding map of weights obtained from the component separation procedure for the foreground model adopted in Ref. (Poletti, Errard et al., in prep.). 
	The average of value of $w_i$ over the pixels is reported in Table~\ref{tab:sigmax_w_res}: we note that the weights are all positive, as they correspond to CMB channels. The maps $m^x_{res,i}$ have been obtained with values of $\sigma_{\Delta x}$ listed in Table~\ref{tab:sigmax_w_res}. These values correspond to the highest standard deviations among the ones in Table~\ref{tab:sigmax} that also satisfy the requirements summarised in Table~\ref{tab:dr_dsyst}.
	
\begin{figure}[!htbp]
		\centering 
		\includegraphics[width=0.5\textwidth]{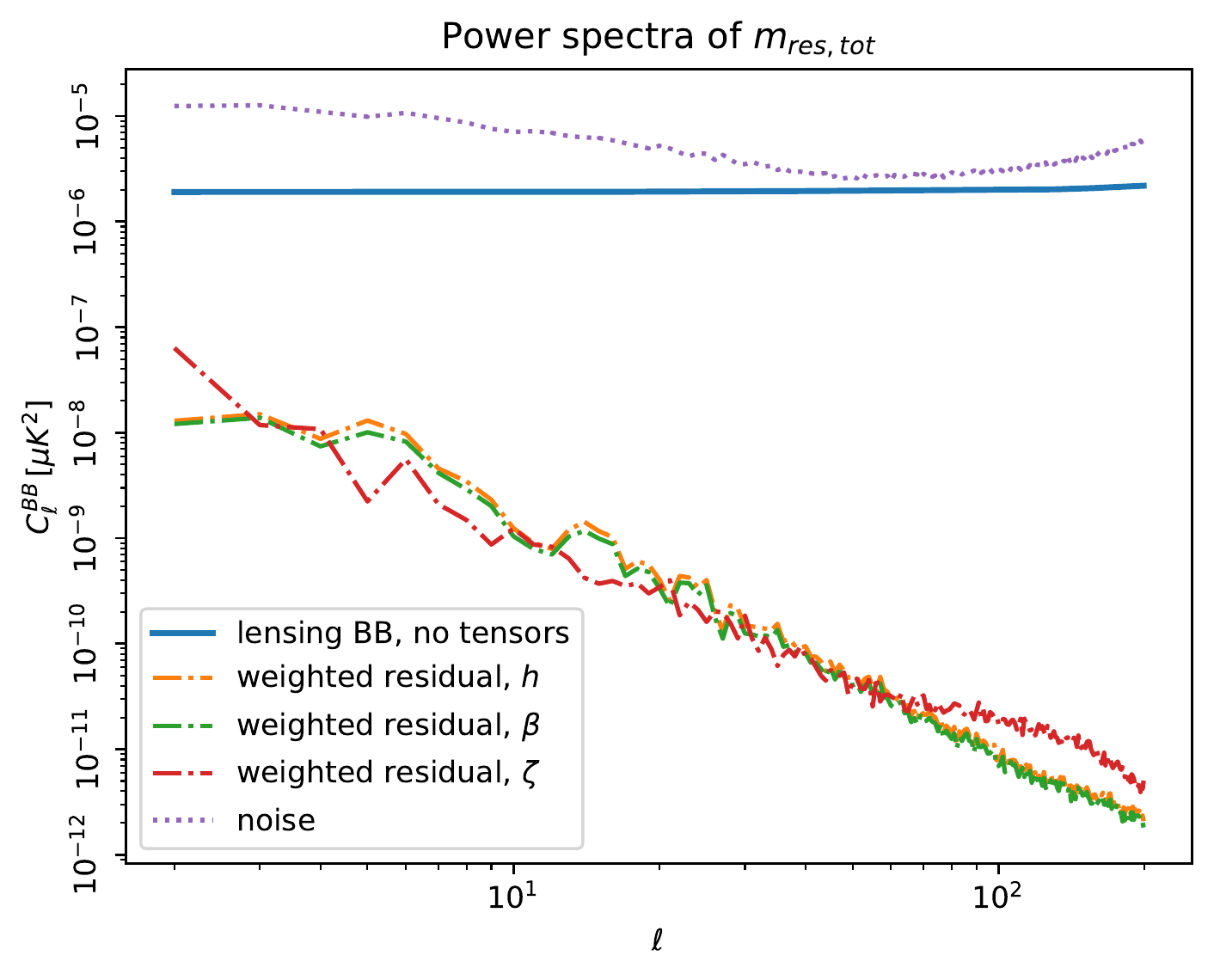}\quad
		\caption{Residual $BB$ power spectra from the coadded residual maps. The thick blue line is the fiducial $C_{\ell}^{BB}$(CMB) with $r = 0$. The dashed-dotted lines are the residual $C_{\ell}^{BB,res}$ of the coadded maps, see Eq. \ref{eq:weight}. Different colors correspond to residual spectra due to a different class of HWP systematic parameter. The dotted line is the noise spectrum $C_{\ell}^{BB,noise}$, as described in Sect.~\ref{subsec:deltar}.} \label{fig:weight_res}
	\end{figure}

	In Fig.~\ref{fig:weight_res}, we show the residual spectra of $ m^x_{res,tot}$.
	These residuals are lower than those obtained individually from each $m^x_{res,i}$. A similar result is obtained in Sect.~\ref{subsec:deltar}. Here, we anticipate that there is a compensation between residual maps of different channels, even though they share the same error realizations on the frequencies where the bands overlap. From this comparison, we argue that the component separation procedure could reduce the residual caused by mismatches in the HWP systematic parameters, at least when the perturbations to the HWP parameters are uncorrelated in frequency (as we assume in this work).
	
	\subsection{Propagation to the tensor-to-scalar ratio} \label{subsec:deltar}
	In the multi-frequency case, we quantify the impact of HWP non-idealities in terms of a possible bias on the estimate of the tensor-to-scalar ratio $r$.
	To evaluate the bias,
	we compute the likelihood of the output spectra:
	\begin{equation} \label{eq:Clres}
		\tilde{C}_{\ell}^\mathrm{BB} = C_{\ell}^{BB,fid} + C_{\ell}^{BB,res} + C_{\ell}^{BB,noise} \,,
	\end{equation}
	
	\noindent where $C^{BB, fid}_{\ell}\equiv C_{\ell}^{BB,lensed}$ is the fiducial $BB$ power spectrum (lensing only, $r=0$), $C_{\ell}^{BB,res}$ is the $BB$ power spectrum of the residual map $\mathcal{R}$ as computed in Sect.~\ref{subsec:depr} and $ C_{\ell}^{BB,noise}$ is the noise spectrum due to foreground residual from component separation and instrumental noise for LiteBIRD (see Poletti, Errard et al., in prep.,
	for details on how this spectrum was obtained). Hereafter, we drop the superscript $BB$ for simplicity. 
	The residual $C_{\ell}^{res}$ is treated as if it were a spurious cosmological signal leading to a bias in the estimate of $r$. 
	
	Next, we adopt the exact likelihood distribution \citep{Gerbino:2019okg,Hamimeche}:
	\begin{equation} \label{eq:3.1}
		\begin{split}
			&-2 \text{ln}\mathcal{\tilde{L}}_i(r) = -2 \text{ln}\mathcal{L}(\tilde{C}_{\ell,i}|C_{\ell}(r)+ C_{\ell}^{noise})  = f_{sky} \sum_{\ell} (2 \ell+1) \left[ \frac{\tilde{C}_{\ell,i}}{C_{\ell}(r)+ C_{\ell}^{noise}} - \text{ln} \left(\frac{\tilde{C}_{\ell,i}}{C_{\ell}(r)+ C_{\ell}^{noise}} \right)  \right] \,,
		\end{split}
	\end{equation}
	
	\noindent where the $i = 1,..,n$ index stands for one specific realization of the $B_i$ solver matrix, $\tilde{C}_\ell$ is the observed power spectrum, and $C_{\ell}(r) = C_{\ell}^{lensed} + C_{\ell}^{tens}(r)$ is the theoretical $BB$ power spectrum for a given value of $r$. The likelihood analysis is restricted to the multipole range $2 \leq \ell \leq 200$ of interest for LiteBIRD~\citep{Hazumi:2021yqq}. The log-likelihoods are then averaged over all the $i$ realizations
	and renormalized to the peak of the distribution. We use a flat prior on $r$. 
	
	
	
	We define $\tilde{p}^\mathrm{res}(r)$ as the posterior distribution corresponding to the likelihood averaged over all the error realizations. The bias on $r$ due to HWP systematics, $\Delta r$, is quantified as the maximum probability value\footnote{This comes from the fact that, in the ideal case of perfect control over systematics, we should recover the fiducial value $r=0$. Therefore, a non-vanishing estimate of $r$ corresponds to a systematics-induced bias in our case.} of the posterior distribution: $\Delta r = r_{peak({\Delta \text{syst.}})}$.
	
	As a final remark, we note that the residual map is computed as the difference with respect to the template map, containing the input CMB map $m_{CMB}$. The latter leaves a distorted signal in the residual map that is in principle dependent on the actual CMB realization. However, this contribution is completely negligible with respect to the foreground-induced residual. For this reason, we can safely neglect the scatter due to the cosmic variance and add the residual power spectrum directly to the fiducial one.

	\subsection{Requirements on the sensitivity for the systematics from the bias $\Delta r$}
\begin{figure*} 
		\centering
		\includegraphics[width= 17 cm]{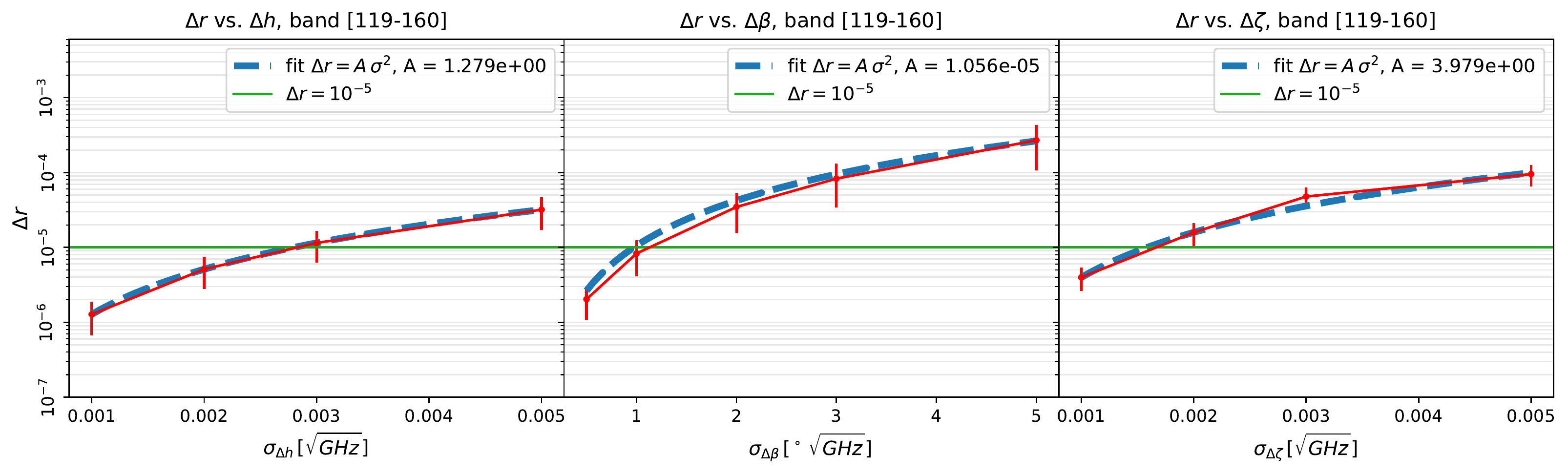}\quad
		\caption{Quadratic fit of the relation $\Delta r \, - \, \sigma_{\Delta x}$, for the band centered at 140 GHz. The green solid line marks the threshold $\Delta r = 10^{-5}$, which we have set to derive a requirement on the highest tolerable $\sigma_{\Delta x}$ when we perturb only one systematic at a time.}\label{fig:dr_s}
	\end{figure*}

	The bias $\Delta r$ is estimated for each value of $\sigma_{\Delta x}$ for each systematic parameter and frequency band. We expect a quadratic relation between the bias on $r$ and the variance of the error realization since $\Delta r \propto C_{\ell} \propto \sigma_{\Delta x}^2$. Indeed, if we plot $\Delta r $ vs. $\sigma_{\Delta x}$, we find that a quadratic fit works well, especially for smaller error variance. In Fig.~\ref{fig:dr_s}, the bias on $r$ due to the perturbation of $\beta$ is larger than the bias coming from the other two parameters ($h,\zeta$). This is due to the fact that we are considering larger $\sigma_{\Delta \beta}$ because of a wider dynamical range of $\beta$ in the HWP model profiles. In the same figure, the error on $\Delta r$ is reported as $\sigma_{\Delta r}/\sqrt{10}$, that is the standard deviation of $\Delta r_i$ from each $i=1,...10$ error realization divided by the square root of the number of realizations.
	We derive the accuracy requirements on each systematic parameter in each band so that $\Delta r \lesssim 10^{-5}$. This threshold is set as 1\% of the expected sensitivity on $r$ from LiteBIRD, that is, $\sigma_r \sim 10^{-3}$. The accuracy requirements are quoted in Table~\ref{tab:dr_dsyst}.

	We also checked the effects of letting two systematic parameters at a time to be perturbed in the analysis. In this case, we find that the bias on $r$ is approximated by the sum of the bias induced individually in the case when one parameter is perturbed, that is, $\Delta r_{x,y} \simeq \Delta r_x + \Delta r_y$, for $x,y = \{h, \beta, \zeta\}$. To allow for a more robust check, we increased the number of error realizations per parameter to 200 for a single band. We first generated maps in which one kind of parameter was perturbed at a time and computed the corresponding averaged $\Delta r_x$. Then we used the same error realizations to generate maps in which two classes of parameters $x,y$ are jointly perturbed. We derived the corresponding $\Delta r_{x,y}$ and compared it against the sum $ \Delta r_x + \Delta r_y$, finding no significant difference. This amounts to having no clear correlation between different parameters within the error $\frac{\sigma_{\Delta r}}{\sqrt{200}}$. This could be due to the fact that we are perturbing the model profiles in a way that is independent both on frequency and on the systematic parameters. A more complex and realistic modeling of the perturbations could change this result. For example, we could allow for some degree of correlation between errors on different systematics, which might be the case if measurements of the HWP parameters were performed at the same time. However, this choice would require a realistic modeling of how measurements are performed, which goes beyond the scope of this paper.

	\begin{table*}[ht]
		\begin{center}
			\caption{Accuracy level required for measurements of HWP parameters $h,\beta,\zeta$ in order to keep the bias on $r$ below $\Delta r \simeq 10^{-5}$}  \label{tab:dr_dsyst}
			\begin{tabular}{ p{2.7cm}|p{3.5cm}|p{3.7cm}|p{3.5cm}  }
				\hline
				\hline
				&\small{$\sigma_{\Delta h}$($\Delta r \simeq 10^{-5}$) [$\sqrt{\text{GHz}}$]} & \small{$\sigma_{\Delta \beta}$($\Delta r \simeq 10^{-5}$) [$^\circ \, \sqrt{\text{GHz}}$]}& \small{$\sigma_{\Delta \zeta}$($\Delta r \simeq 10^{-5}$) [$\sqrt{\text{GHz}}$]}\\
				\hline
				\hline
				100 GHz (LFT)& $\leq 0.0029$ & $\leq 3.3$  &  $\leq 0.0016$ \\
				\hline
				100 GHz (MFT)& $\leq 0.0030$  & $\leq 2.7$ & $\leq 0.0017$  \\
				119 GHz (MFT)& $\leq 0.0041$ & $\leq 2.1$ &      $\leq 0.0015$    \\
				140 GHz  (MFT)& $\leq 0.0028$ & $\leq 1.1$ &    $\leq 0.0016$   \\
				166 GHz (MFT)& $\leq 0.0018$ & $\leq 1.4$ & $\leq 0.0013$ \\
				total (MFT) & $\leq 0.0018$  & $\leq 1.1$  & $\leq 0.0013$  \\
				\hline
				195 GHz  (HFT)& $\leq 0.0017$ & $\leq 1.1$ & $\leq 0.0010$ \\
				\hline
				\hline
			\end{tabular}
		\end{center}
		{\raggedright \textbf{Notes.} Threshold values are given for individual LiteBIRD MFT frequency bands and one band for LFT and HFT (quoted with their band center). The total MFT threshold is set by the lowest threshold in MFT bands. The error variance $\sigma_{\Delta x}$ is quoted per frequency resolution. \par}
	\end{table*}
	
	Finally, we estimate the bias on $r$ caused by the weighted average of residual maps from each channel (Eq.~\ref{eq:weight}). We obtain a $\Delta r^x_{tot}$ which is always smaller than the weighted average $\Delta r^x_{weight} = \frac{\sum_{j=1}^{\mathrm{N_{pixels}}}  (\sum_i w^2_i \Delta r^x_i/\sum_i w^2_i )}{\mathrm{N_{pixels}}} $ of the $\Delta r^x_i$ corresponding to each $m^x_{res,i}$ (in Table~\ref{tab:sigmax_w_res}):
	\begin{equation}
		\begin{split}
			\Delta r^h_{tot} = 1.3 \times 10^{-6} \quad &< \quad \Delta r^h_{weight} = 3.9 \times 10^{-6}\\
			\Delta r^{\beta}_{tot} = 1.2 \times 10^{-6} \quad &< \quad \Delta r^{\beta}_{weight} = 8.3 \times 10^{-6}\\
			\Delta r^{\zeta}_{tot} = 2.6 \times 10^{-6} \quad &< \quad \Delta r^{\zeta}_{weight} = 5.9 \times 10^{-6} .\\
		\end{split}
	\end{equation}
	We weight the biases with $w^2_i$, as $\Delta r \propto C_{\ell}$, which is quadratic in the map. 
	It is possible that a non-parametric component separation procedure would relax the requirements shown in Table~\ref{tab:dr_dsyst}.
	
	\begin{table*}
		\begin{center} 
			\caption{Average weights, $\bar{w}_i$, assigned to each frequency channel from component separation and requirements on the accuracy, $\sigma_{\Delta x}$, needed to measure specific classes of HWP non-ideal properties $x\equiv h,\,\zeta,\,\beta$.
			}  \label{tab:sigmax_w_res}
			\begin{tabular}{ p{1.9 cm}|p{1.8cm}|p{1.8cm}|p{1.8cm}|p{1.8cm}|p{1.8cm}|p{1.8cm}}
				\hline
				\hline
				& 100 LFT &100 MFT&119 MFT&140 MFT&166 MFT&195 MFT \\
				\hline
				\hline
				$\bar{w}_i$ & 0.043 & 0.064 & 0.179  &  0.156 &  0.206 & 0.053 \\
				\hline
				\hline          
				$\sigma_h$ \small{$[\sqrt{\text{GHz}}]$} & 0.003&0.003&0.003&0.002&0.001&0.001 \\ 
				\hline
				$\Delta r_h$ & $1.12 \times 10^{-5}$ &  $1.02 \times 10^{-5}$&  $7.16 \times 10^{-6}$&  $2.98 \times 10^{-6}$ & $1.13 \times 10^{-6}$ &  $3.43 \times 10^{-6}$ \\
				\hline
				\hline
				$\sigma_{\beta}$ \small{$[^\circ\sqrt{\text{GHz}}]$} &  3&2&2&1&1&1\\ 
				\hline
				$\Delta r_{\beta}$ & $8.06 \times 10^{-6}$ &  $6.60 \times 10^{-6}$&  $1.83 \times 10^{-5}$&  $2.27 \times 10^{-6}$ & $4.50 \times 10^{-6}$ &  $7.79 \times 10^{-6}$ \\
				\hline
				\hline
				$\sigma_{\zeta}$ \small{$[\sqrt{\text{GHz}}]$} & 0.001 &  0.001 &  0.001 &  0.001 &  0.001 &  0.001 \\ 
				\hline
				$\Delta r_{\zeta}$ & $3.82 \times 10^{-6}$ &  $1.72 \times 10^{-6}$&  $4.34 \times 10^{-6}$&  $6.95 \times 10^{-6}$ & $6.36 \times 10^{-6}$ &  $1.70 \times 10^{-5}$ \\
				\hline
				\hline
			\end{tabular}
		\end{center}
		{\raggedright \textbf{Notes.} These requirements amount to setting the bias $\Delta r$ on the estimate of the tensor-to-scalar ratio, $r,$ to below 1\% the expected sensitivity from LiteBIRD, that is, $\Delta r \lesssim 10^{-5}$.         
			\par}
	\end{table*}
	
	\section{Conclusions} \label{sec:conclusion}
	
	In this work, we study the impact of non idealities of the half-wave plate (HWP) in the context of future cosmic microwave background (CMB) observations, focusing on the case of a LiteBIRD-like satellite mission. We consider the following classes of non-idealities: departure from unitary transmission ($h$), spurious phase shift ($\beta$), and cross-polarization (mixing of orthogonal polarization components, $\zeta$). Any mismatch between the measured properties of the HWP and the actual properties that enter in the construction of the time-ordered data (TOD) during observations can propagate throughout the analysis pipeline and bias the final science products down to the tensor-to-scalar ratio, $r,$ estimate. We have first presented at length the formalism describing how light propagation is affected by a non-ideal rotating HWP.  We have developed an agile simulation suite to quickly reproduce the LiteBIRD scanning strategy and find the relative on-the-fly mapmaking solution, with little computational cost. To do so, we considered a simplified scenario where light propagates with normal incidence through the optical system (including the HWP) and is collected by a single pair of polarization-sensitive detectors at boresight. Because of this simplified setting, we are not able to capture some additional systematic effects, such as the HWP synchronous signal, which has been observed in different experiments~\citep{Ritacco:2016due,Johnson:2006jk,Kusaka:2013pla}.
	The optical system has been described in the Jones formalism and we have also shown the conversion to the Mueller formalism. The full expressions of the Mueller matrix elements of the non-ideal (not rotating) HWP are presented in Appendix~\ref{app:fullM}. 
	
	We first focused on the case of an input CMB-only sky observed at a single frequency. This is motivated by the fact that we wanted to single out the effects of HWP non-idealities on the reconstructed CMB spectra while neglecting any other source of contamination (e.g., color-correction due to bandpass integration). We have shown results obtained in two scenarios: a) in the case in which a mismatch persists between the HWP parameters entering the TOD and the ones used in the map-making solution; b) in the case in which the two set of parameters are identical, albeit they are still non-ideal. As expected, our results show that scenario b) minimizes the propagation of HWP-induced systematic effects to CMB spectra.
	
	We then moved on to a more realistic study with a frequency-dependent input signal (including also foregrounds) modulated by a frequency-dependent HWP profile. We have considered the four MFT frequency bands of LiteBIRD centered at [100, 119, 140, 166] GHz, and the closest LFT/HFT bands to the CMB channels (centered at 100/195 GHz; see Table~\ref{tab:bands}). We assumed a top-hat bandpass profile, for simplicity. In this multi-frequency study, we have only focused on the case in which the profile of the TOD HWP does not match the profile of the mapmaking HWP. We adopted simulated frequency profiles for the departure from unitary transmission, $h,$ and the non-ideal phase shift $\beta$ in the MFT frequency bands, provided by finite-element simulations of the MHFT LiteBIRD MHWPs. The waveplate designs are based on previous developments and realizations~\citep{pisano2020development}.
	The profiles for $\zeta$ are always fixed to a realistic \citep{Pisano:crosspolcite} level of 0.01 in all the bands, as this parameter was not included in the suite of simulations at our disposal. In the LFT (also not included in the simulation suite) and HFT bands, we used constant profiles also for $h$ and $\beta$. To simulate a mismatch between the TOD HWP and the map-making HWP, all the profiles were perturbed with Gaussian-distributed errors, uncorrelated both in frequency and among the different parameters. We noted that this simple procedure allow our results to be basically independent from the initial shape in frequency. This justifies our choice a posteriori of fixing the parameters for the LFT/HFT bands, as well as the value of $\zeta$ in each channel, to a constant value. 
	
	In this multi-frequency study, a template map obtained from the observation of the same input sky (CMB and foregrounds) with an ideal HWP was deprojected from the realistic output maps to obtain maps of residuals. The template has been generated with exactly the same foreground model adopted for the input maps. This is equivalent to assuming a perfect knowledge of the foreground sky. Of course, this may not be the case with actual observations -- however, it is justified by the need to not include uncertainty on the foreground modeling on top of the effect of the systematics in the residual maps. Our assumption guarantees that by construction, the residual maps vanish when the TOD HWP parameters perfectly match those associated with the mapmaking. 
	
	In the multi-frequency case, the CMB spectra have been extracted from the residual maps after applying a galactic mask ($f_\mathrm{sky}=70\%$). These residual spectra, in addition to foreground residual from component separation and instrumental noise for LiteBIRD, have been fed to an exact likelihood to quantify their induced bias $\Delta r$ on $r$ with respect to the fiducial estimate obtained in absence of systematics residuals. The bias has been quantified for each class of systematic effects individually and in each individual frequency band. By imposing that $\Delta r \leq 10^{-5}$ (1\% of the expected sensitivity on $r$ from LiteBIRD), we set a requirement on the accuracy needed on each HWP parameter in each band.
	
	We repeated the analysis by allowing for pairs of non-ideal parameters to be perturbed simultaneously to check for correlated effects between classes of non-idealities. We found that the bias $\Delta r$ from a joint variation is consistent with the sum of the biases corresponding to perturbing each of the two parameters at a time, with the same error. This is enough to exclude significant correlations between non-ideal HWP parameters given our experimental setup. In fact, our assumption that Gaussian perturbations fully capture the error in the measurement of HWP parameters is likely to be unrealistic. For example, it is possible to have errors that are correlated within the same frequency band. In addition, errors on different parameters might be correlated if their measurements are simultaneous. To implement this kind of perturbation scheme, we would need a realistic model of how measurements of HWP properties are performed. We defer this study to a future work.
	
	We also provide the results of the coaddition of residual maps from the different frequency channels. We find a general reduction of $\Delta r$ for the final coadded map. This could point to the fact that a component separation procedure might mitigate the impact of HWP non-idealities thanks to a cancellation among the frequency channels -- at least in our setup.
	
	Nonetheless, allowing for a correction of these systematic effects in the mapmaking process remains key to mitigating their impact on science products. We showed that considering an ideal HWP in the map-making procedure could lead to $\Delta r \approx \mathcal{O}(10^{-3}-10^{-2})$, depending on the amplitude of each systematic parameter. Furthermore, some calibration procedures could be attempted for the parameters $h$ and $\beta$, which mediate the polarization efficiency (see Eqs.~\ref{eq:Muellerx_h}, \ref{eq:Muellerx_b}). In Appendix~\ref{app:polang}, we showed that $\zeta$ behaves similarly to a rotation of the polarization angle (see Appendix~\ref{app:polang}), and could be thus reabsorbed in the calibration of the latter. Some complications could arise from the frequency dependence of those parameters, however.
	When this work was in preparation, a study by A. Duivenvoordeen and collaborators (\citeyear{Duivenvoorden:2020xzm}) 
	was published on the same topic, and we would like to point out that our analysis nicely complements the findings of those authors. Here, we offer a pedagogical approach to the use of a HWP in CMB experiments. We also provide a thorough discussion of the complementarity between the Jones and Mueller formalisms in the context of CMB polarimetry. Finally, we highlight a significant, and possibly problematic, effect: the fact that the in-band variation of the properties of a non-ideal HWP can affect the observed signal and the reconstructed sky maps by introducing an effective band integration that depends also on the HWP rotation angle (see Sect.~\ref{subsec:depr}). This effect can potentially lead to a direction-dependent bandpass mismatch. 

	\section*{Acknowledgment}
	
	We acknowledge the use of \texttt{numpy} \citep{harris2020array}, \texttt{matplotlib} \citep{Hunter:2007} , \texttt{healpy} \citep{Zonca2019}, \texttt{pysm} \citep{Thorne_2017} and \texttt{pymaster} \citep{Alonso_2019} software packages, and the use of computing resources at CINECA. SG, MG, LP, AG, ML, PN acknowledge the financial support from the INFN InDark project and from the COSMOS network (www.cosmosnet.it) through the ASI (Italian Space Agency) Grants 2016-24-H.0 and 2016-24-H.1-2018. JE acknowledges the French
	National Research Agency (ANR) grants ANR-B3DCMB (ANR-17-CE23-0002) and ANR-BxB
	(ANR-17-CE31-0022).
	
	\bibliographystyle{aa}
	\bibliography{HWP_systematics_aa,Planck_bib}
	
	
	
	\appendix
	\section{Full expression of Mueller matrix elements for a non-ideal HWP}\label{app:fullM}
	In the following, we recall that $h_i$, $\zeta_i$, $\chi_i$ for $i=1,2$, and $\beta$ are the parameters used to describe the deviations from the ideal behavior of a HWP. In particular, $A_i$ and $B_i$ for $i=1,2$ are the elements of the corresponding Jones matrix. Even if $A_1$ is real, we treat it as complex in the most general expression of each Mueller matrix element. The diagonal blocks of the Mueller matrix given in Eq.~\ref{eq:mueller}, for $\theta = 0$, are:
	\begin{equation} \label{eq:2.5} 
		\begin{split}
			T_1 &= \frac{1}{2} \Big((1 + h_1)^2 + (1 + h_2)^2 + \zeta_1^{2} + \zeta_2^{2} \Big) = \frac12 \Big( A_1^* A_1 + A_2^* A_2 + B_1^* B_1 + B_2^* B_2  \Big)\\
			T_2 &= \frac{1}{2} \Big((1 + h_1)^2 + (1 + h_2)^2 - \zeta_1^{2} - \zeta_2^{2}\Big)  = \frac12 \Big( A_1^* A_1 + A_2^* A_2 - B_1^* B_1 - B_2^* B_2  \Big)\\
			\rho_1 &= \frac{1}{2}  \Big((1 + h_1)^2 - (1 + h_2)^2 - \zeta_1^2 + \zeta_2^2 \Big)  = \frac12 \Big( A_1^* A_1 - A_2^* A_2 - B_1^* B_1 + B_2^* B_2  \Big)\\
			\rho_2 &= \frac{1}{2}  \Big((1 + h_1)^2 - (1 + h_2)^2 + \zeta_1^2 - \zeta_2^2 \Big)  = \frac12 \Big( A_1^* A_1 - A_2^* A_2 + B_1^* B_1 - B_2^* B_2  \Big)\\
			c_1 & = -(1 + h_1) (1 + h_2) \, \text{cos}(\beta) + \zeta_1 \zeta_2 \, \text{cos}(\chi_1-\chi_2) = \text{Re}[-(1 + h_1) (1 + h_2)e^{i \beta} + \zeta_1 e^{i \chi_1}(\zeta_2 e^{i \chi_2})^*] = \\
			& = \text{Re}[A_1^* A_2 + B_1 B_2^*] \\
			c_2 & = -(1 + h_1) (1 + h_2) \, \text{cos}(\beta) - \zeta_1 \zeta_2 \, \text{cos}(\chi_1-\chi_2) = \text{Re}[-(1 + h_1) (1 + h_2)e^{i \beta} - \zeta_1 e^{i \chi_1}(\zeta_2 e^{i \chi_2})^*]\\
			&  = \text{Re}[A_1^* A_2 - B_1 B_2^*]\\
			s_1 & = -(1 + h_1) (1 + h_2) \, \text{sin}(\beta) + \zeta_1 \zeta_2 \, \text{sin}(\chi_1-\chi_2) = \text{Im}[-(1 + h_1) (1 + h_2)e^{i \beta} + \zeta_1 e^{i \chi_1}(\zeta_2 e^{i \chi_2})^*] =\\
			&  = \text{Im}[A_1^* A_2 + B_1 B_2^*]\\
			s_2 & = -(1 + h_1) (1 + h_2) \, \text{sin}(\beta) - \zeta_1 \zeta_2 \, \text{sin}(\chi_1-\chi_2) = \text{Im}[-(1 + h_1) (1 + h_2)e^{i \beta} - \zeta_1 e^{i \chi_1}(\zeta_2 e^{i \chi_2})^*] = \\
			& = \text{Im}[A_1^* A_2 - B_1 B_2^*] .
		\end{split}
	\end{equation}

	For the off-diagonal blocks we have:
	\begin{equation} \label{eq:2.6} 
		\begin{split}
			a_1 & = (1 + h_1) \, \zeta_1 \, \text{cos}(\chi_1) - (1+h_2) \, \zeta_2 \, \text{cos}(\beta-\chi_2) = \text{Re}[(1 + h_1) \, \zeta_1 e^{i \chi_1} -(1 + h_2)e^{i \beta} ( \zeta_2 e^{i \chi_2})^*] = \\
			& = \text{Re}[A_1^* B_1 + A_2 B_2^*] \\
			a_2 & = (1 + h_1) \, \zeta_1 \, \text{cos}(\chi_1) + (1+h_2) \, \zeta_2 \, \text{cos}(\beta-\chi_2) = \text{Re}[(1 + h_1) \, \zeta_1 e^{i \chi_1} +(1 + h_2)e^{i \beta} ( \zeta_2 e^{i \chi_2})^*] = \\
			& = \text{Re}[A_1^* B_1 - A_2 B_2^*] \\
			a_3 & = (1 + h_1) \, \zeta_2 \, \text{cos}(\chi_2) - (1+h_2) \, \zeta_1 \, \text{cos}(\beta-\chi_1) = \text{Re}[(1 + h_1) \, \zeta_2 e^{i \chi_2} -(1 + h_2)e^{i \beta} ( \zeta_1 e^{i \chi_1})^*] =\\
			& = \text{Re}[A_1^* B_2 + A_2 B_1^*] \\
			a_4 & = (1 + h_1) \, \zeta_2 \, \text{cos}(\chi_2) + (1+h_2) \, \zeta_1 \, \text{cos}(\beta-\chi_1) = \text{Re}[(1 + h_1) \, \zeta_2 e^{i \chi_2} +(1 + h_2)e^{i \beta} ( \zeta_1 e^{i \chi_1})^*] =\\
			& = \text{Re}[A_1^* B_2 - A_2 B_1^*]\\
			b_1 & = -(1 + h_1) \, \zeta_1 \, \text{sin}(\chi_1) + (1+h_2) \, \zeta_2 \, \text{sin}(\beta-\chi_2) = \text{Im}[(1 + h_1) \, (\zeta_1 e^{i \chi_1})^* -((1 + h_2)e^{i \beta})^*  \zeta_2 e^{i \chi_2}] = \\
			& =  \text{Im}[A_1 B_1^* + A_2^* B_2]   \\
			b_2 & = -(1 + h_1) \, \zeta_1 \, \text{sin}(\chi_1) - (1+h_2) \, \zeta_2 \, \text{sin}(\beta-\chi_2) = \text{Im}[(1 + h_1) \, (\zeta_1 e^{i \chi_1})^* + ((1 + h_2)e^{i \beta})^*  \zeta_2 e^{i \chi_2}] = \\
			& = \text{Im}[A_1 B_1^* - A_2^* B_2] \\
			b_3 & = (1 + h_1) \, \zeta_2 \, \text{sin}(\chi_2) - (1+h_2) \, \zeta_1 \, \text{sin}(\beta-\chi_1) = \text{Im}[(1 + h_1) \, \zeta_2 e^{i \chi_2} -(1 + h_2)e^{i \beta} ( \zeta_1 e^{i \chi_1})^*] =\\
			&  = \text{Im}[A_1^* B_2 + A_2 B_1^*]\\
			b_4 & = (1 + h_1) \, \zeta_2 \, \text{sin}(\chi_2) + (1+h_2) \, \zeta_1 \, \text{sin}(\beta-\chi_1) = \text{Im}[(1 + h_1) \, \zeta_2 e^{i \chi_2} +(1 + h_2)e^{i \beta} ( \zeta_1 e^{i \chi_1})^*] = \\
			& = \text{Im}[A_1^* B_2 - A_2 B_1^*] \,. 
		\end{split}
	\end{equation}
	
	\noindent
	It can be noted that if $\zeta_1 = \zeta_2 = 0$ (no cross-polarization), the off-diagonal blocks with $a_{1,2,3,4}$ and $b_{1,2,3,4}$ would be zero and $T_1 = T_2 = T$, $\rho_1 = \rho_2 = \rho$, $c_1 = c_2 = c$ and $s_1 = s_2 = s$. These expressions agree with results that could be found in the literature, such as~\citep{Bryan2010}.
	
	\begin{figure}[!htbp]
		\centering
		\includegraphics[width=0.7\textwidth]{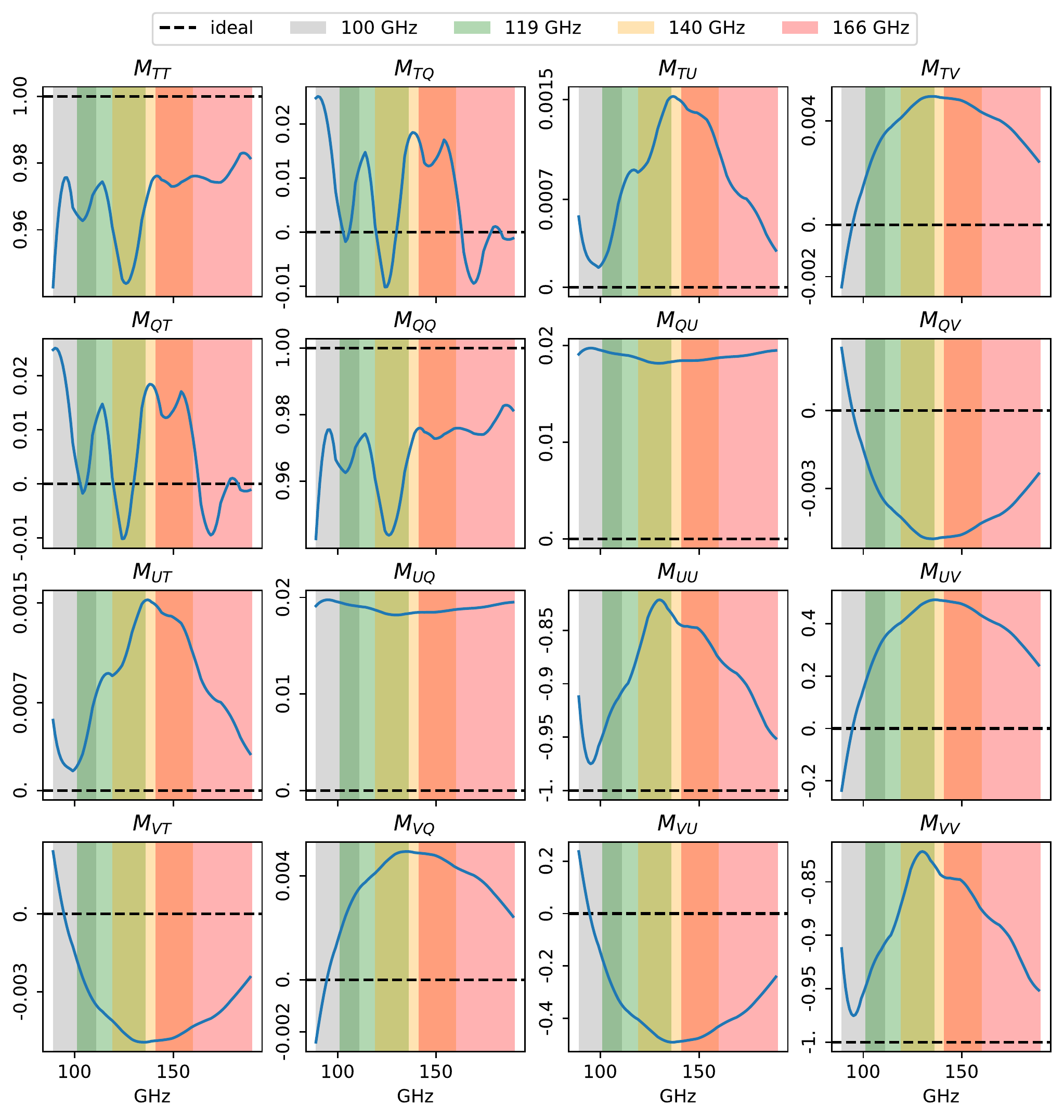}
		\caption{Graphical representation of the Mueller matrix of a non-ideal HWP. Each panel corresponds to the profile of a matrix element as a function of frequency. We used the simulated profiles of $h$ and $\beta$ for the MFT bands, while we fix $\zeta_{1,2} = 0.01$ and $\chi_{1,2} = 0$. The black dashed line represents the case of an ideal HWP, the shaded vertical bands correspond to the four MFT bands, labeled by their central value. The non-diagonal blocks do not vanish as a result of $\zeta_{1,2} \neq 0$ (see Eq.~\ref{eq:2.6}).} \label{fig:prof_M}
	\end{figure}
	
	\subsection{Rotating HWP followed by a polarizer}
	Now we are able to present the case of the whole optical elements $M_i = M_{pol,i}M_\mathrm{rot}^{T}M_\mathrm{HWP}M_\mathrm{rot}$, where $i = x,y$, focusing on the elements entering the bolometer equation ($M^{TT}_{i}, M^{TQ}_{i}, M^{TU}_{i}, M^{TV}_{i}$). We note that the rotation matrix, $M_{\psi}$, which accounts for the position angle $\psi$ of the telescope in the sky frame and have to precede $M_i$, simply leads to the substitution $\theta \rightarrow \theta + \frac{\psi}{2}$ in the following expressions. If we expand at first order in $h,\zeta_{1,2},\chi_{1,2}$, the matrix elements of $M_x$, we find:
	
	\begin{equation} \label{eq:Muellerx}
		\begin{split}
			M^{TT}_{x} &=  \frac{1}{2} (|J_{11}|^2 + |J_{12}|^2) \simeq \\
			& \simeq \frac12  \Big(1+h_1+h_2+(h_1-h_2)\cos(2\theta) +\left(\zeta_1 \cos\chi_1 \cos\beta -\zeta_2 \cos\chi_2 \right)\sin(2\theta) \Big) \\ 
			M^{TQ}_{x} &=  \frac{1}{2} (|J_{11}|^2 - |J_{12}|^2) \simeq \\
			& \simeq \frac14 \left(1+h_1+h_2 \right) \left(1-\cos\beta\right)+ \frac12 (h_1-h_2)\cos(2\theta) + \frac14 \left(1+h_1+h_2 \right) \left(1+\cos\beta\right)\cos(4\theta)-\\ 
			&-\frac12 \left(\zeta_1  \cos\chi_1-\zeta_2  \cos\chi_2 \cos\beta\right)\sin (2\theta)-\frac14\left(\zeta_1  \cos\chi_1 +\zeta_2  \cos\chi_2 \right) \left(1+\cos\beta\right)\sin (4\theta)\\
			M^{TU}_{x} &= \text{Re}[(J_{11} J_{12}^*)] \simeq \\
			& \simeq \frac14\left(\zeta_1 \cos\chi_1-\zeta_2 \cos\chi_2\right) \left(1-\cos\beta\right)+\frac12 (h_1-h_2)\sin(2\theta) + \frac14\left(1+h_1+h_2 \right) \left(1+\cos\beta\right)\sin(4\theta)\\ 
			&+\frac12 \left(\zeta_1 \cos\chi_1 -\zeta_2 \cos\chi_2 \cos\beta\right)\cos (2\theta)+\frac14 \left(\zeta_1 \cos\chi_1 +\zeta_2 \cos\chi_2\right) \left(1+\cos\beta\right)\cos (4\theta)\\
			M^{TV}_{x} &= \text{Im}[(J_{11} J_{12}^*)] \simeq -\frac12 \sin\beta \sin(2 \theta) \,.
		\end{split}
	\end{equation}
	
	\noindent
	The corresponding elements when the polarizer is along the $y$ direction are: 
	
	\begin{equation} \label{eq:Muellery}
		\begin{split}
			M^{TT}_{y} &=  \frac{1}{2} (|J_{21}|^2 + |J_{22}|^2) \simeq \\
			& \simeq \frac12  \Big(1+h_1+h_2 -(h_1-h_2)\cos(2\theta) -\left(\zeta_1 \cos\chi_1 \cos\beta -\zeta_2 \cos\chi_2 \right)\sin(2\theta) \Big) \\ 
			M^{TQ}_{y} &=  \frac{1}{2} (|J_{21}|^2 - |J_{22}|^2) \simeq \\
			&  \simeq - \frac14 \left(1+h_1+h_2 \right) \left(1-\cos\beta\right)+ \frac12 (h_1-h_2)\cos(2\theta) - \frac14 \left(1+h_1+h_2 \right) \left(1+\cos\beta\right)\cos(4\theta)-\\ 
			&-\frac12 \left(\zeta_1  \cos\chi_1-\zeta_2  \cos\chi_2 \cos\beta\right)\sin (2\theta)+\frac14\left(\zeta_1  \cos\chi_1 +\zeta_2  \cos\chi_2 \right) \left(1+\cos\beta\right)\sin (4\theta)\\
			M^{TU}_{y} &= \text{Re}[(J_{21}^{*}J_{22})] \simeq \\
			& \simeq - \frac14\left(\zeta_1 \cos\chi_1-\zeta_2 \cos\chi_2\right) \left(1-\cos\beta\right)+\frac12 (h_1-h_2)\sin(2\theta) - \frac14\left(1+h_1+h_2 \right) \left(1+\cos\beta\right)\sin(4\theta) \\ 
			&+\frac12 \left(\zeta_1 \cos\chi_1 -\zeta_2 \cos\chi_2 \cos\beta\right)\cos (2\theta) - \frac14 \left(\zeta_1 \cos\chi_1 +\zeta_2 \cos\chi_2\right) \left(1+\cos\beta\right)\cos (4\theta)\\
			M^{TV}_{y} &= \text{Im}[(J_{21}^{*}J_{22})] \simeq + \frac12 \sin\beta \sin(2 \theta) \,.
		\end{split}
	\end{equation}
	
	\noindent
	It is clear that $M^{TT}_{y}(\theta) = M^{TT}_{x}(\theta + \frac{\pi}{2})$, $M^{TQ}_{y}(\theta) = -M^{TQ}_{x}(\theta + \frac{\pi}{2})$, $M^{TU}_{y}(\theta) = -M^{TU}_{x}(\theta + \frac{\pi}{2})$ and $M^{TV}_{y}(\theta) = M^{TV}_{x}(\theta + \frac{\pi}{2})$.

	\section{Relation between $\zeta$ and a rotation of the polarization angle} \label{app:polang}
\begin{figure}[htbp!]
		\centering
		\includegraphics[width=0.7\textwidth]{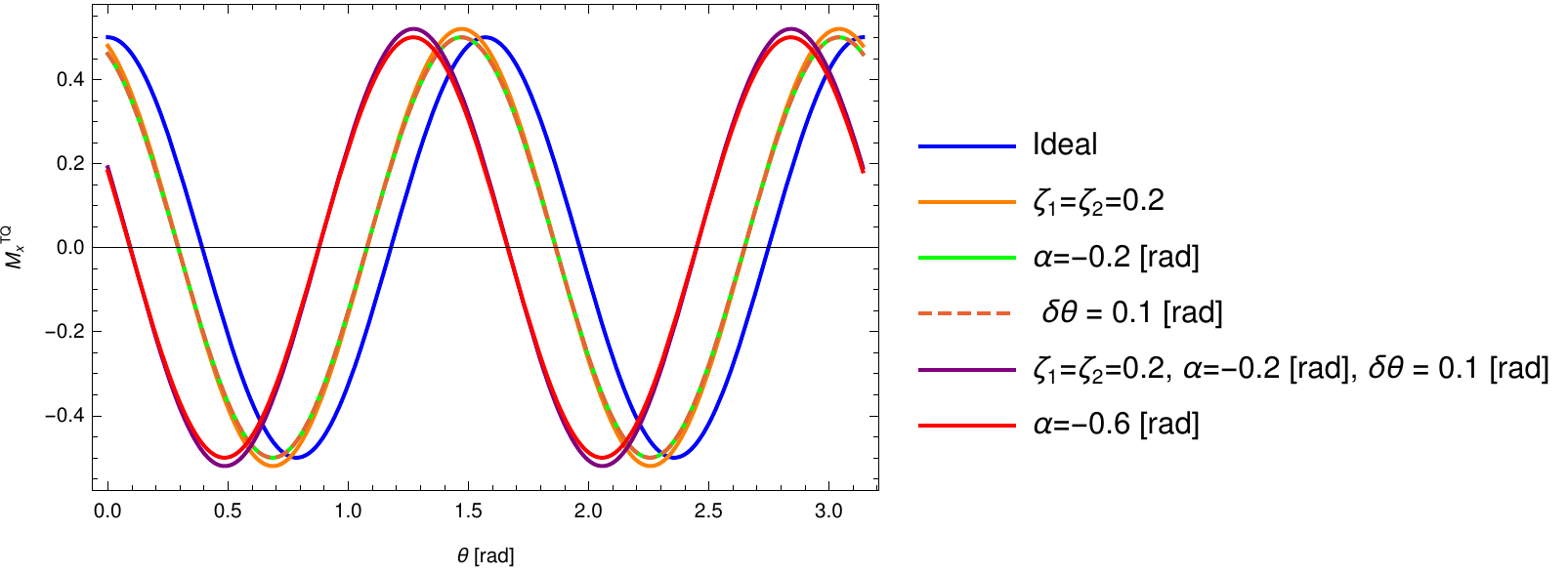}
		\caption{Mueller matrix element $M^{TQ}_x(\theta)$ that modulates the Stokes-$Q$ component of the sky signal, shown as a function of the HWP rotation angle $\theta$. We assume an ideal HWP. In blue, we show the ideal case of vanishing polarization angle and vanishing uncertainty on the HWP rotation angle, $\alpha, \delta \theta = 0$ [rad]. In solid orange, we show the case of $\zeta_1 = \zeta_2 = 0.2$ and $ \alpha,\delta \theta = 0$ [rad]. In green, the ideal HWP with $\delta \theta = 0, \alpha = -0.2$ [rad]. In dashed orange, we have the ideal HWP with $\delta \theta = 0.1, \alpha = 0$ [rad]. We note that the orange dashed and the green lines overlap perfectly. They also partly overlap with the solid orange line, albeit the latter shows a slightly different amplitude. The purple line shows all the effects combined together. Finally, the red line corresponds to $\alpha = 3 \times (-0.2)$ [rad]. We can see that the shift of the red curve is equivalent to three times the effects described with the yellow, green, and dashed curves.}\label{fig:all}
	\end{figure}

	We consider the effect of both the cross-polarization parameter $\zeta$ and a miscalibration in the polarization angle. The parameter $\zeta$ is defined in Eq.~\ref{eq:realistic} and is responsible for the mixing of orthogonal polarizations.
	The Mueller matrix of a rotating HWP\footnote{The rotation matrix is defined as a clockwise rotation}, followed by a polarization-sensitive detector along the $x$ direction is given by:
	\begin{equation}
		M_x(h,\beta,\zeta,\theta)\equiv M_{pol,x}M_\mathrm{rot}^{T}(\theta)M_\mathrm{HWP}(h,\beta,\zeta)M_\mathrm{rot}(\theta).
	\end{equation}
	A miscalibration of the polarization angle can be modeled as an additional rotation by an angle $\alpha$ on the focal plane, such that:
	\begin{equation}
		M_x(h,\beta,\zeta,\theta,\alpha)\equiv M_{pol,x}M_\mathrm{rot}(\alpha)M_\mathrm{rot}^{T}(\theta)M_\mathrm{HWP}(h ,\beta,\zeta )M_\mathrm{rot}(\theta).
	\end{equation}
	We emphasize the effect of $\zeta$ and $\alpha$ one at a time, setting all the other non-ideal parameters to zero. Considering first $\zeta_{1,2}$ only and expanding $M^{TQ/U}_x(\zeta)$ at first order in $\zeta_{1,2}$:
	\begin{equation} \label{eq:Mqu_zeta}
		\begin{split}
			M^{TQ}(\zeta) & \simeq \frac12 \cos(4 \theta) -\frac12 \left(\zeta_1  -\zeta_2   \right)\sin (2\theta)-\frac12\left(\zeta_1  +\zeta_2 \right) \sin (4\theta),\\
			M^{TU}(\zeta) & \simeq \frac12 \sin(4 \theta) + \frac12 \left(\zeta_1  -\zeta_2 \right)\cos (2\theta)+\frac12 \left(\zeta_1 +\zeta_2 \right) \cos (4\theta) .
		\end{split}
	\end{equation}
	Instead, considering only $\alpha \neq 0$:
	\begin{equation}
		\begin{split}
			M^{TQ}(\alpha) &= \frac12\cos(2 \alpha) \cos(4 \theta) +\frac12 \sin(2 \alpha) \sin(4 \theta),\\
			M^{TU}(\alpha) &=\frac12 \cos(2 \alpha) \sin(4 \theta) - \frac12 \sin(2 \alpha) \cos(4 \theta) 
		\end{split}
	\end{equation}
	and expanding at first order in $\alpha$:
	\begin{equation} \label{eq:Mqu_alfa}
		\begin{split}
			M^{TQ}(\alpha) &\simeq \frac12 \cos(4 \theta) +  \alpha \sin(4 \theta),\\
			M^{TU}(\alpha) &\simeq \frac12 \sin(4 \theta) -  \alpha \cos(4 \theta) .
		\end{split}
	\end{equation}
	For small $\alpha$, Eqs.~\ref{eq:Mqu_zeta},~\ref{eq:Mqu_alfa} describe a similar effect as long as $\alpha \simeq -\frac12 (\zeta_1+\zeta_2)$. The only difference between the two equations is the presence of $2 \theta$ terms in the case of $\zeta$. However, we expect them to be averaged out by the LiteBIRD scanning strategy (see main text for discussion).
	In Fig.~\ref{fig:all}, we show $M^{TQ}_x, M^{TU}_x$ as a function of the HWP rotation angle $\theta$. We can see that the modification with respect to the ideal case induced by both $\zeta, \alpha \neq 0$ is similar. However, we can appreciate that $\zeta$, contrarily to $\alpha$, also affects the amplitude of the curves, as a non-ideal $J_\mathrm{HWP}$ is not an orthogonal matrix.
	
	
	We can also see that the effect of $\alpha$ is equivalent to that of an uncertainty in the HWP rotation angle:
	\begin{equation}
		\begin{aligned}
			M_x(\theta + \delta \theta) =& M_{pol,x} M_\mathrm{rot}^{T}(\theta+ \delta \theta)M_\mathrm{HWP}(h = 0,\beta = 0,\zeta = 0)M_\mathrm{rot}(\theta+ \delta \theta) =\\ 
			=& M_{pol,x}M_\mathrm{rot}(-2 \delta \theta)M_\mathrm{rot}^{T}(\theta)M_\mathrm{HWP}(h = 0,\beta = 0,\zeta = 0)M_\mathrm{rot}(\theta)
		\end{aligned}
	\end{equation}
	\noindent
	Provided $\alpha = -2 \delta \theta$, the two effects are equivalent. From the discussion above, we see that if $\delta \theta = \frac14 (\zeta_1+\zeta_2)$, the uncertainty in the HWP rotation angle is first-order equivalent to the effect of $\zeta$.
	The effects are in principle additive, as seen in Fig.~\ref{fig:all}.
	
	As a final remark, we want to stress that both $\zeta$ and $\alpha$ can  generally serve as the frequency-dependent parameters.

	\section{Impact on $\Delta r$ of $h$ and $\beta$ frequency profiles} \label{app:flat_prof}
\begin{figure}[htbp!]
		\centering
		\includegraphics[width=0.8\textwidth]{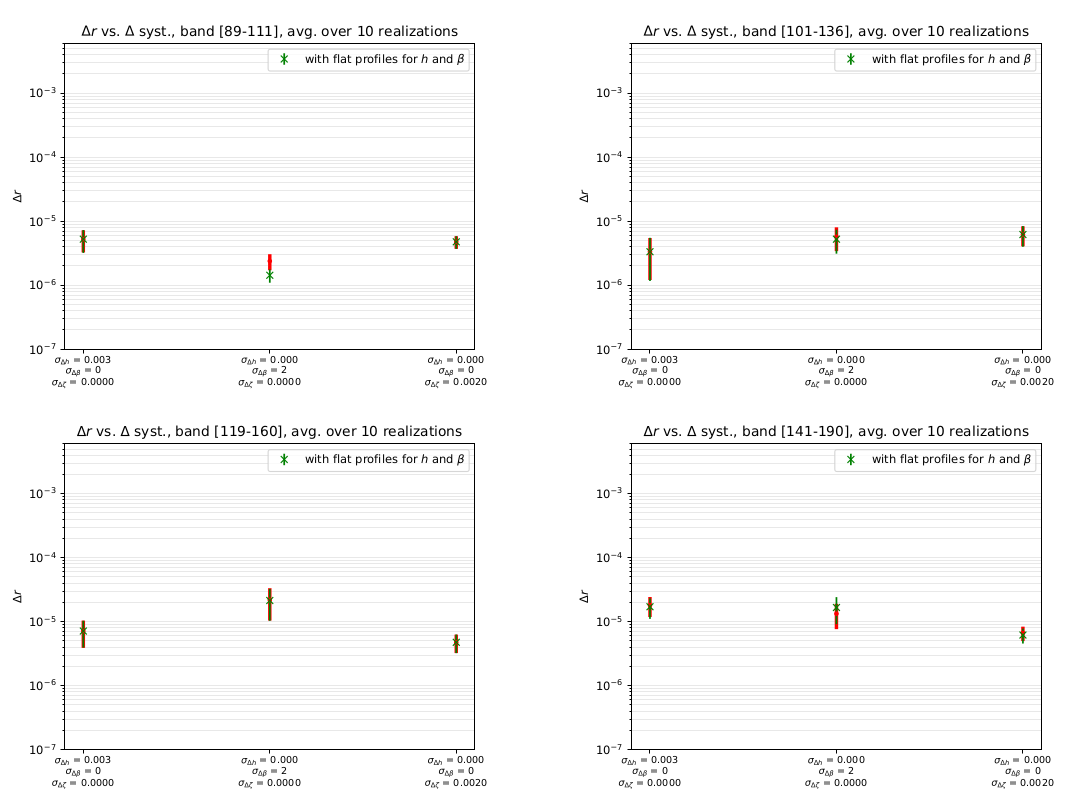}
		\caption{Comparison between the $\Delta r$ obtained with constant parameters and those obtained with the simulated profiles, in four MFT bands. In each case, only one systematics $X \in \{h,\beta,\zeta\}$ is perturbed with the $\sigma_X$ indicated in the label. The red points refer to the case with simulated profiles, the green crosses to the case with constant ones.}\label{fig:flat_prof}
	\end{figure}

	We checked  how much the shape of the frequency profiles of the non-ideal parameters impacts the estimated $\Delta r$. We generated ten simulations for each of the selected MFT band, perturbing either $h, \beta,$ or $\zeta$ with one of the $\sigma$ listed in Table \ref{tab:sigmax}; in addition,  instead of using the simulated profiles for $h$ and $\beta$ (Figures \ref{fig:prof_h} and \ref{fig:prof_b}), we fixed both to be constant in frequency. The constant value for $h$ and $\beta$ in each band is fixed to their average value in the band. The error realizations are the same used with the simulated profiles (see Sect. \ref{subsec:setup}). In all the cases, $\zeta$ is constant and equal to 0.01. We then derived the average $\Delta r$ and its standard deviation from the ten simulations for each realization and compared them to the ones obtained with $h$ and $\beta$ varying in frequency (see Sect. \ref{subsec:deltar}). We found that in all the cases, the obtained $\Delta r$s are compatible within the errors. The largest discrepancy, more than 1$\sigma$, is found when perturbing $\beta$ in the 100 GHz band (see Fig. \ref{fig:flat_prof}). It is worth noticing that this result assumes uncorrelated errors in the band, in accordance with all the analyses presented in this paper. 
	
\end{document}

%% file: Planck.tex
\def\setsymbol#1#2{\expandafter\def\csname #1\endcsname{#2}}
\def\getsymbol#1{\csname #1\endcsname}

\newbox\tablebox    \newdimen\tablewidth
\def\leaderfil{\leaders\hbox to 5pt{\hss.\hss}\hfil}
\def\tablenote#1 #2\par{\begingroup \parindent=0.8em
    \abovedisplayshortskip=0pt\belowdisplayshortskip=0pt
    \noindent
    $$\hss\vbox{\hsize\tablewidth \hangindent=\parindent \hangafter=1 \noindent
    \hbox to \parindent{$^#1$\hss}\strut#2\strut\par}\hss$$
    \endgroup}

\def\L2{\ifmmode L_2\else $L_2$\fi}

\def\DeltaT{\ifmmode \Delta T\else $\Delta T$\fi}
\def\deltat{\ifmmode \Delta t\else $\Delta t$\fi}
\def\fknee{\ifmmode f_{\rm knee}\else $f_{\rm knee}$\fi}
\def\Fmax{\ifmmode F_{\rm max}\else $F_{\rm max}$\fi}
\def\solar{\ifmmode{\rm M}_{\mathord\odot}\else${\rm M}_{\mathord\odot}$\fi}
\def\Msolar{\ifmmode{\rm M}_{\mathord\odot}\else${\rm M}_{\mathord\odot}$\fi}
\def\Lsolar{\ifmmode{\rm L}_{\mathord\odot}\else${\rm L}_{\mathord\odot}$\fi}
\def\inv{\ifmmode^{-1}\else$^{-1}$\fi}
\def\mo{\ifmmode^{-1}\else$^{-1}$\fi}
\def\sup#1{\ifmmode ^{\rm #1}\else $^{\rm #1}$\fi}
\def\expo#1{\ifmmode \times 10^{#1}\else $\times 10^{#1}$\fi}
\def\,{\thinspace}
\def\lsim{\mathrel{\raise .4ex\hbox{\rlap{$<$}\lower 1.2ex\hbox{$\sim$}}}}
\def\gsim{\mathrel{\raise .4ex\hbox{\rlap{$>$}\lower 1.2ex\hbox{$\sim$}}}}

\def\simprop{\mathrel{\raise .4ex\hbox{\rlap{$\propto$}\lower 1.2ex\hbox{$\sim$}}}}
\def\deg{\ifmmode^\circ\else$^\circ$\fi}
\def\pdeg{\ifmmode $\setbox0=\hbox{$^{\circ}$}\rlap{\hskip.11\wd0 .}$^{\circ}
          \else \setbox0=\hbox{$^{\circ}$}\rlap{\hskip.11\wd0 .}$^{\circ}$\fi}
\def\arcs{\ifmmode {^{\scriptstyle\prime\prime}}
          \else $^{\scriptstyle\prime\prime}$\fi}
\def\arcm{\ifmmode {^{\scriptstyle\prime}}
          \else $^{\scriptstyle\prime}$\fi}
\newdimen\sa  \newdimen\sb
\def\parcs{\sa=.07em \sb=.03em
     \ifmmode \hbox{\rlap{.}}^{\scriptstyle\prime\kern -\sb\prime}\hbox{\kern -\sa}
     \else \rlap{.}$^{\scriptstyle\prime\kern -\sb\prime}$\kern -\sa\fi}
\def\parcm{\sa=.08em \sb=.03em
     \ifmmode \hbox{\rlap{.}\kern\sa}^{\scriptstyle\prime}\hbox{\kern-\sb}
     \else \rlap{.}\kern\sa$^{\scriptstyle\prime}$\kern-\sb\fi}
\def\ra[#1 #2 #3.#4]{#1\sup{h}#2\sup{m}#3\sup{s}\llap.#4}
\def\dec[#1 #2 #3.#4]{#1\deg#2\arcm#3\arcs\llap.#4}
\def\deco[#1 #2 #3]{#1\deg#2\arcm#3\arcs}
\def\rra[#1 #2]{#1\sup{h}#2\sup{m}}

\def\dots{\relax\ifmmode \ldots\else $\ldots$\fi}
\def\WHzsr{\ifmmode $W\,Hz\mo\,sr\mo$\else W\,Hz\mo\,sr\mo\fi}
\def\mHz{\ifmmode $\,mHz$\else \,mHz\fi}
\def\GHz{\ifmmode $\,GHz$\else \,GHz\fi}
\def\mKs{\ifmmode $\,mK\,s$^{1/2}\else \,mK\,s$^{1/2}$\fi}
\def\muKs{\ifmmode \,\mu$K\,s$^{1/2}\else \,$\mu$K\,s$^{1/2}$\fi}
\def\muKRJs{\ifmmode \,\mu$K$_{\rm RJ}$\,s$^{1/2}\else \,$\mu$K$_{\rm RJ}$\,s$^{1/2}$\fi}
\def\muKHz{\ifmmode \,\mu$K\,Hz$^{-1/2}\else \,$\mu$K\,Hz$^{-1/2}$\fi}
\def\MJysr{\ifmmode \,$MJy\,sr\mo$\else \,MJy\,sr\mo\fi}
\def\MJysrmK{\ifmmode \,$MJy\,sr\mo$\,mK$_{\rm CMB}\mo\else \,MJy\,sr\mo\,mK$_{\rm CMB}\mo$\fi}
\def\microns{\ifmmode \,\mu$m$\else \,$\mu$m\fi}

\def\muK{\ifmmode \,\mu$K$\else \,$\mu$\hbox{K}\fi}
\def\microK{\ifmmode \,\mu$K$\else \,$\mu$\hbox{K}\fi}
\def\muW{\ifmmode \,\mu$W$\else \,$\mu$\hbox{W}\fi}
\def\kms{\ifmmode $\,km\,s$^{-1}\else \,km\,s$^{-1}$\fi}
\def\kmsMpc{\ifmmode $\,\kms\,Mpc\mo$\else \,\kms\,Mpc\mo\fi}
\providecommand{\sorthelp}[1]{}

%% file: shortcuts.tex
\def\NHUNIT{\ifmmode {\rm \,cm^{-2}} \else $\rm \,cm^{-2}$ \fi} 

\def\aap{{A\&A}}

\def\apjs{{ApJ Supp.}}
\def\muKcmb{\ifmmode \,\mu$K$_{\rm CMB}$\else \,$\mu$K$_{\rm CMB}$\fi}
\newcommand{\OmegaM}{\ifmmode\Omega_{\rm M}\else $\Omega_{\rm M}$\fi}

\providecommand{\text}[1]{\rm{#1}}

\providecommand{\muK}{\mu\rm{K}}

\newcommand{\begm}{\begin{pmatrix}}
\newcommand{\enm}{\end{pmatrix}}

\def\pmb#1{\setbox0=\hbox{#1}%
    \kern-.025em\copy0\kern-\wd0
    \kern.05em\copy0\kern-\wd0
    \kern-.025em\raise.0433em\box0}

\def\p2Y{\;_2Y}
\def\m2Y{\;_{-2}Y}
\def\beglet{
  \addtocounter{equation}{1}%
  \setcounter{parentequation}{\value{equation}}%
  \setcounter{equation}{0}%
  \def\theequation{\arabic{parentequation}\alph{equation}}%
  \ignorespaces
}
\def\endlet{
  \setcounter{equation}{\value{parentequation}}%
  \def\theequation{\arabic{equation}}%
}
\providecommand{\beglet}{\begin{subequations}}
\providecommand{\endlet}{\end{subequations}}

\newcommand{\mksym}[1]{\ifmmode {\rm #1}\else #1\fi}

\providecommand{\text}[1]{\rm{#1}}

\providecommand{\muK}{\mu\rm{K}}

\newcommand\ba{\begin{eqnarray}}
\newcommand\ea{\end{eqnarray}}
\newcommand\bea{\begin{eqnarray}}
\newcommand\eea{\end{eqnarray}}

\newcommand\be{\begin{equation}}
\newcommand\ee{\end{equation}}

